\pdfoutput=1
\documentclass{aa}  
\usepackage{longtable}
\usepackage{siunitx}
\usepackage{graphicx}
\usepackage{txfonts}
\usepackage{tabularx}
\usepackage{subcaption}
\usepackage{lscape}
\usepackage{placeins}
\usepackage{hyperref}  
\hypersetup{
    colorlinks=true,
    allcolors=blue
}

\usepackage{natbib}
\bibpunct{(}{)}{;}{a}{}{,}
\begin{document}

   \title{Precise radial velocities of giant stars}

   \subtitle{XVIII. Rotation periods based on photometry, activity, and radial velocity}

   \author{Silas Janke\inst{1}\corrauth{silas.janke@stud.uni-heidelberg.de}
        \and Sabine Reffert\inst{1}\email{sreffert@lsw.uni-heidelberg.de}
        }

   \institute{Landessternwarte, Zentrum für Astronomie der Universität Heidelberg, Königstuhl 12, 69117 Heidelberg, Germany}

   \date{Received October 7, 2026; accepted November 11, 2026}
 
\abstract
{Precise radial velocities were collected over 12 years for a sample of 373 presumably single G- and K-type giants at Lick Observatory. Observations of a large fraction of these stars continued with SONG, CARMENES, and Waltz, covering a total of 27 years to date. Some display radial velocity variations that cannot clearly be attributed to substellar companions.}
{We aim to find rotation periods of these stars to avoid confusion of spot-induced radial velocity modulation with a planetary signal and to characterise these potential exoplanet host stars.}
{We searched for periodic variations in archival activity data from Mount Wilson Observatory, \textsc{Hipparcos} epoch photometry, and the Lick radial velocity data, using generalised Lomb--Scargle periodograms and Gaussian process modelling. To constrain possible rotation periods, we used upper limits derived from rotational velocities, $v \sin i$, and stellar radii.}
{We found plausible rotation periods for a total of 31 stars from the sample, ranging from about 25\,d to 800\,d. Of these, 18 show a consistent period in at least two independent datasets. Eleven of these include a compatible period in the radial velocity data, supporting the notion that rotational modulation contributes to their observed radial velocity variations. Stars with shorter rotation periods generally show higher levels of chromospheric activity.}
{The identified rotation periods provide a basis for distinguishing spot-induced radial velocity modulation from planetary signals in the Lick giant sample, while also contributing to our understanding of rotation in single G and K giants.}


\maketitle
\nolinenumbers

\section{Introduction}
\label{sec:Intro}

In the 31 years since the first discovery of an exoplanet around a main-sequence star \citep{mayor1995}, the quality of observations and, thus, the number and diversity of discovered exoplanets have vastly improved. While the boundaries of instrumental limitations have been pushed to improve sensitivity towards lower-mass planets, the presence of intrinsic stellar activity can complicate the retrieval of low-amplitude planetary signals or even mimic such a signal in Doppler data \citep{zhao2022}. It is thus crucial to characterise and understand stellar intrinsic activity, rotation, and oscillations to properly disentangle them from the genuine signatures of planetary companions in precise radial velocity (RV) datasets.

G and K giant stars offer several advantages as targets for RV planet searches. Their expanded atmospheres produce narrow spectral lines enabling precise Doppler measurements, they are bright and abundant, and as intermediate-mass stars at a later evolutionary stage they allow the study of the interplay between planetary and stellar evolution. The first planet orbiting a giant star was discovered around $\iota$ Draconis in 2002 with the RV method \citep{frink2002}. Today, a total of 151 substellar companions (planets or brown dwarfs) are known to orbit giant stars, almost all of them discovered from RV variations\footnote{\url{https://www.lsw.uni-heidelberg.de/users/sreffert/giantplanets/giantplanets.php}}. 
As stars evolve off the main sequence onto the giant branch, their expanding envelopes cause them to spin down significantly through conservation of angular momentum, resulting in typically low chromospheric and photometric activity levels compared to their main-sequence counterparts \citep{skumanich1972, gray1989, pasquini2000, barnes2003}. Nevertheless, the RV signals of evolved stars can still be subject to the influence of stellar activity. Their main complication is a significant intrinsic RV jitter from solar-like oscillations, which reaches 10--20\,m\,s$^{-1}$ for K giants \citep{frandsen2002}. Since oscillation periods range from hours to a few days -- much shorter than the orbital periods of planetary companions -- this contribution can be mitigated by observational strategies such as multiple consecutive observations per epoch \citep{kjeldsen1995}. Other reasons for RV variations include surface granulation \citep{tayar2019}, non-radial oscillations \citep{spaeth2024}, or surface features modulated by the rotation of the star around its own axis \citep{hatzes1999}. This rotational modulation is the main focus of this work.

Due to rotational Doppler broadening, spectral lines have a positive and negative RV part, corresponding to the parts of the star that rotate towards and away from the observer. When inhomogeneous surface features, such as dark spots or bright plages, are present, there is an imbalance between the negative and positive RV part of the spectral line, manifesting in a residual RV shift. Rotation moves these features across the stellar disc, which can result in a periodic RV variation at the rotation period and possibly its harmonics \citep{saar1997, boisse2011}. Disentangling these variations from planetary signals, especially of lower-mass planets around active stars, requires an understanding of the star's activity from an intensive analysis of the spectral data. Additional monitoring, for example through photometry or the observation of activity indicators such as H$\alpha$ or Ca\,\textsc{ii} H and K lines, can be another valuable source to improve the understanding of a potential planet host star \citep{vanderburg2016}.
Spectroscopic line-profile diagnostics derived directly from the RV spectra are a further powerful tool: the chromatic index and the differential line width \citep{zechmeister2018} have proven particularly valuable for M~dwarfs \citep{lafarga2021}. They have also been applied in the context of giant stars from this survey \citep{spaeth2025}.

Rotation periods of single G and K giants have been measured in a number of earlier studies. \citet{choi1995} derived periods for several bright giants from long-term monitoring of Ca\,\textsc{ii} H and K emission using the Mount Wilson HK Project (HKP) dataset. \citet{schroder2020} used the same dataset to identify rotation periods of four K giants in the Hyades cluster and study their evolutionary stages. \citet{lehtinen2020} measured rotation periods for 92 subgiants and giants in the same dataset and showed that, in terms of the Rossby number, they follow the rotation--activity relation of main-sequence stars.
Photometric surveys have also been used to identify surface rotation periods of giant stars. 
\citet{fekel1986} identified a small group of single G--K giants with moderately high rotation rates and chromospheric activity levels, hypothesising that they may have evolved from rapidly rotating early-type stars.
After taking photometric observations over one year, \citet{strassmeier1999} attributed most of the variability periods they found in their sample of chromospherically active stars to rotational spot modulation. 
\citet{henry2000} found low-amplitude periodic variability in about one fourth of their G and K giant sample, but attributed most of them to (non-)radial pulsations, while proposing rotational modulation only for a few stars with high chromospheric activity.
This high activity level was tied to magnetic field strength when \citet{auriere2015} detected magnetic fields via Zeeman signatures in a sample of active single G--K giants and demonstrated a tight correlation between field strength and rotation period, confirming a dynamo origin.
Using data from the Kepler satellite, \citet{ceillier2017} found photometric rotation periods for 361 red giants, combined with asteroseismic mass measurements. They found that only rapidly rotating giants ($P<160$\,d) show detectable spot modulation, since slow rotators tend to be magnetically inactive, and attributed the rapid rotation to speed-up events such as binary interactions.
Despite this body of work, a systematic search for rotation periods combining photometric, activity, and RV data across a large, uniformly observed sample of single G and K giants remains lacking.

The goal of this study is to improve the understanding of RV variations in G and K giants. We focus on stars from the Lick sample, for which we have extensive RV data. We combine the RVs with archival photometric and activity data, aiming to find periodic variations that are consistent with rotation periods.
We describe the sample, data, period search, and spot models in Sects.~\ref{sec:sample}--\ref{sec:spotmodel}. We present the periods, their relation to other parameters, and individual stars in Sects.~\ref{sec:rotation_periods} and \ref{sec:individual_stars}. We discuss and summarise our findings in Sects.~\ref{sec:discussion} and \ref{sec:summary}.

\section{Stellar sample and parameters}
\label{sec:sample}

The sample being studied in this work consists of 373 G and K giants that were observed with the Hamilton Spectrograph at Lick Observatory from 1999 to 2011 \citep{frink2001}. The stars were initially selected from the \textsc{Hipparcos} catalogue \citep{perryman1997} to have no stellar companions and a photometric variability below 0.01\,mag in the \textsc{Hipparcos} photometry. For the detailed selection criteria, we refer to \citet{frink2001} and \citet{reffert2015}. The median of the magnitude scatter (field H46 in the \textsc{Hipparcos} catalogue) is 0.005\,mag and the maximum is 0.07\,mag. All stars are brighter than $V = 6$\,mag, placing them at or above the saturation limit of many large-scale photometric surveys.

The Lick survey led to the detection of several substellar companions (planets or brown dwarfs) around these stars \citep[e.g.][]{frink2002, trifonov2014, quirrenbach2019}. While the RV variations of some stars were attributed to planets with high certainty, especially the very luminous and large evolved stars show RV signals that are still to be conclusively explained. Possible explanations include non-radial oscillations \citep{spaeth2024}, but there are still investigations ongoing into the nature of RV variations of these stars \citep{heeren2021, spaeth2025}. Additional independent datasets can support efforts to understand their RV variations and to characterise the stars.

The stellar parameters for the sample were derived by \citet{stock2018} using Bayesian inference on a grid of stellar evolutionary models. They used two priors -- one for the initial mass function and one taking into account the evolutionary timescales at each position in the colour--magnitude diagram (CMD). The stellar radii, $R_*$, from the most probable evolutionary state for all stars with at least one significant period are listed in Table~\ref{tab:results}. Additionally, measurements of the rotational velocity, $v \sin i$, were collected from different sources (shown in the same table). From these values, the projected rotation period can be calculated as

\begin{equation}
P_\mathrm{max} = \frac{2 \pi R_*}{v \sin i}.
\end{equation}

As the inclination is unknown, this gives a nominal upper limit on the stellar rotation period. Since the errors in $R_*$ are asymmetric, we used Monte Carlo sampling to estimate $P_\mathrm{max}$. We drew $10^5$ samples from $R_*$ and $v\sin i$, assuming (split-)Gaussian errors. We calculated the resulting $P_\mathrm{max}$. The values and errors in Table~\ref{tab:results} correspond to the 16th, 50th, and 84th percentiles of these samples. A conservative upper limit was obtained by adopting the $1\,\sigma$ upper bound. In Sect.~\ref{sec:rotation_periods}, we describe how we used them to check whether a measured period is plausible to be a rotation period. Their distribution can be seen in Fig.~\ref{fig:p_rot_hist}. In slowly rotating giant stars, the line broadening due to macroturbulence can be comparable to or even larger than rotational broadening \citep{gray2013}. Furthermore, the linewidth nears typical spectral resolutions as absorption lines become very narrow. This makes the uncertainties in the determination of $v \sin i$ quite large, often more than 20\,\%. Nevertheless, it can provide a good estimate to check the plausibility of a photometric rotation period.

\section{Observations}
\label{sec:obs}
\subsection{Lick RV data}

The sample stars were monitored with the 0.6\,m Coudé Auxiliary Telescope (CAT) at UCO/Lick Observatory using the Hamilton Echelle Spectrograph, at a theoretical resolution of $R\approx60\,000$ \citep{vogt1987}. The observations spanned the time frame from 1999 to 2011. An RV precision of 5--8\,m\,s$^{-1}$ was reached \citep{reffert2015}, using the iodine cell technique \citep{butler1996} in a wavelength range of 5000--5800\,\AA. The RV data used in this work were obtained using the standard Lick RV reduction pipeline. On average, there are about 40 observations per star.

\subsection{\textsc{Hipparcos} epoch photometry}

The \textsc{Hipparcos} satellite was launched in 1989 with the primary goal of conducting astrometric measurements of around 120\,000 stars \citep{perryman1997}. Since the satellite visited all stars many times during the mission to measure changes in their positions, photometric data were obtained at each visit, resulting in the \textsc{Hipparcos} epoch photometry annex \citep{vanleeuwen1997}. The individual flux measurements contain quality flags. We discarded data points with a flag value greater than 7, since this indicates that several problems occurred during the observation, making the measurement less reliable. This threshold was adopted from \citet{koen2002}.

Besides the magnitude scatter used for the sample selection, the catalogue gives a photometric variability class: one sample star (HIP 87933; Sect.~\ref{sec:xi_her}) is a periodic variable, 30 are microvariables (possibly variable with an amplitude $<0.03$\,mag), and 14 unsolved variables. Using the signal-to-noise ratio of periodic signals instead of the original criteria, \citet{koen2002} found new periodic variables, with 12 of them in the Lick sample, so the \textsc{Hipparcos} data may contain more periods.

\subsection{HK Project \texorpdfstring{Ca\,\textsc{ii}}{CaII} data}

The HK Project (HKP) at Mount Wilson Observatory was started in 1966 by Olin Wilson \citep{wilson1968} and continued until 2002 by several others. Measurements of Ca\,\textsc{ii} line intensity were carried out to study stellar chromospheric activity, quantified by the Mount Wilson S index. It is a dimensionless measure for non-radiative chromospheric heating caused by magnetic activity and defined as 

\begin{equation}
    S = \alpha \: \frac{H + K}{V + R},
\end{equation}

where $\alpha$ is an instrumental constant. H and K are the fluxes in triangular passbands with a nominal width of 1\,\AA{} (full width at half maximum of 1.09\,\AA{}), centred on the H and K lines of Ca\,\textsc{ii} at 396.8\,nm and 393.4\,nm, respectively. V and R are the continuum fluxes in two 2\,nm wide rectangular passbands centred at 390.1\,nm and 400.1\,nm \citep{baliunas1998}. While the project initially aimed at main-sequence stars, it was extended to evolved stars starting in 1983. For these, the H and K passbands were widened to 2\,\AA{} because of the wider emission cores in the Ca\,\textsc{ii} H and K lines of giant stars \citep{duncan1991}. 

A higher S index is associated with active regions on a star, mainly bright plages in the chromosphere \citep{cretignier2023a} and, to a smaller extent, starspots in the photosphere \citep{sowmya2023}. When active regions are distributed unevenly in longitude, rotation causes them to move in and out of sight, leading to a modulation of the S index at the rotation period or its harmonics. The possibility of finding rotation periods from S index variations has been proven before, for main-sequence stars \citep[see][]{baliunas1983, gilliland1985} as well as giant stars \citep{choi1995, schroder2020}.

A total of 221 stars from our sample are included in the HKP dataset published in 2016, spanning the years 1966--1995 \citep{radick2018}; for our stars, the observations begin in 1980. Many of these stars were observed only sparsely, but 57 stars were observed on at least 20 nights, and 22 of them on at least 100 nights, 17 of which are spread over more than a decade, allowing us to study variability in their chromospheric emission. Only observations starting from 1983 were used in the analysis. The few earlier measurements of our stars were taken with the 1\,\AA{} H and K passbands and lie below the later measurements of the same stars, so they cannot be placed on the same scale.

\begin{figure}[ht]
   \centering
   \includegraphics[width=\linewidth]{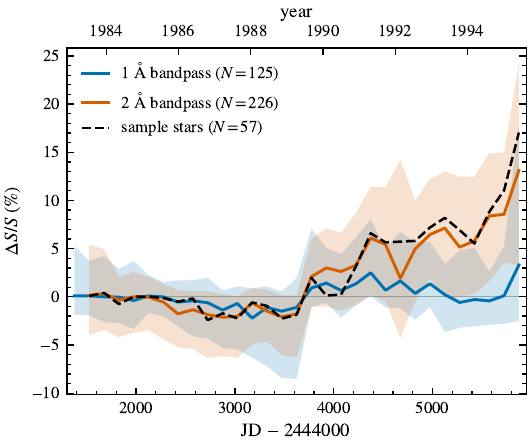}
      \caption{Long-term drift of the HK Project S index. Each archive star's S index is binned into 150\,d intervals and expressed as a percentage of that star's own median level, so that stars of different activity can be combined and the median over many stars retains only what they have in common. The solid curves show the median over all stars observed with the 1\,\AA{} (blue) and 2\,\AA{} (orange) bandpass and the shaded bands show the 16th--84th percentile spread between stars. The dashed curve shows the median of the 57 sample stars with at least 20 nights of data, all of which were observed with the 2\,\AA{} setting.}
         \label{fig:hkp_all}
   \end{figure}

The observations mostly contain three data points per night, taken in rapid succession. Each observation contains a weight, $w$, which gives the photon-limited precision of the measurement as $\sigma_S/S = 1/\sqrt{w}$. We binned the data into nightly data points, since the exposures of a single night are not independent measurements and would otherwise lead to an overestimated significance of periodic signals. Each nightly data point is the mean of its $n$ exposures, placed at their mean epoch. For each night, we calculated the standard error of the mean, floored by the photon-limited precision of the nightly mean,
\begin{equation}
    \sigma_\mathrm{night} = \max\left(s,\, \tilde{\sigma}_S\right)/\sqrt{n},
\end{equation}
where $s$ is the standard deviation of the exposures and $\tilde{\sigma}_S$ the median of their photon-limited uncertainties. The floor is necessary because a standard error estimated from only three exposures is itself highly uncertain. When the exposures happen to agree closely, that night would get a disproportionate weight in the later analysis.

The S index of our sample stars carries a long-term drift that is shared across the archive (see Fig.~\ref{fig:hkp_all}). The drift depends on the bandpass the observation was made with, which is recorded in the instrument code of every measurement. It is flat to within 2\,\% until late 1989, then steps up and rises steadily until the end of the archive in 1995 in the 2\,\AA{} data, while stars measured in the 1\,\AA{} setting stay flat. Our sample was observed entirely with the 2\,\AA{} setting, and ends on average about 17\,\% higher than at the start of observations.

In some stars, there is an indication of a long-term activity trend on top of the instrumental trend. In cases where this trend shows a periodic structure, we report its period as a possible activity cycle (Table~\ref{tab:results}). These periods were taken from the highest periodogram peak of the S index prior to detrending, searched only at periods longer than the detrending cutoff, $P_{\rm cut}$ (see below). They range from about 4 to 15\,yr. Since this series still contains the instrumental drift, the cycle lengths should be regarded as indicative only. The periods around 12\,yr could be analogues to the solar activity cycle, but the baseline of the HKP dataset is still not long enough to confirm that the signals are indeed periodic.
The rotation periods we expect for our sample stars are generally much shorter than the timescale of both the instrumental and activity long-term trends. Thus, we applied a detrending to prevent low-frequency power and spurious periods from dominating the periodograms. Because the amplitude and shape of the instrumental drift vary from star to star, a single global correction would over-correct some stars and under-correct others. Rather than trying to disentangle instrumental and physical trends, we therefore detrended each star individually, removing both at once. The raw S index data and the detrending fits are shown together with the periodograms in the plots in the supplementary material.
The trend was modelled as a cosine series, $m(t) = a_0 + \sum_{k=1}^{K} a_k \cos(\pi k (t-t_0)/T)$, where $T$ is the observed span and $K = \lfloor 2T/P_{\rm cut} \rfloor$. The $k$-th term has period $2T/k$, so the model contains only periods longer than $P_{\rm cut}$ by construction: subtracting it is a high-pass filter that follows a drift of any shape, yet cannot absorb the much shorter rotation signals.
The coefficients were obtained by least squares, with the curvature of the model penalised inside observing gaps so that the fit crosses them on a straight line instead of swinging between seasons, and with three iterations of $3\sigma$ outlier rejection, so that individual deviant nights cannot bend the trend. The rejected points were excluded from the fit only and remained in the series.
The strongest signal below $0.8\,P_{\rm cut}$ was fitted simultaneously with the trend but not subtracted, so that the trend function cannot take up power it shares with the rotation signal on gappy sampling. We used $P_{\rm cut} = 1000$\,d, raised to between 1500 and 4000\,d for stars whose variability is slower, and left the stars undetrended that did not show significant long-period power.

\section{Period search}
\label{sec:periodsearch}

\subsection{GLS periodogram}

We used the \texttt{PyPeriod} implementation of the generalised Lomb--Scargle (GLS) periodogram \citep{zechmeister2009}. The GLS periodogram can be described as a least-squares fit of sine waves on a regular frequency grid. For each frequency, a power is returned that indicates how well it represents the data. We searched for periods between 40\,d and $10^4$\,d with an oversampling factor of 20.
We used the normalisation from \citet{zechmeister2009}, whereby a power of 1 can be interpreted as a perfect sine fit. The best-fitting period was taken from the highest-power peak and its uncertainty was estimated from the curvature of the peak, using the function provided in the package.
To assess the significance of a peak, we calculated the false alarm probability (FAP) from 2000 random permutations of the data. In each permutation, the detrended nightly values were shuffled together with their uncertainties and searched with the same periodogram settings as the data. The 95th, 99th, and 99.9th percentiles of the highest peak across the permutations define the power levels corresponding to FAPs of 5, 1, and 0.1\,\%. We used a significance threshold of 5\,\%, but flagged periods with $0.1\,\%<\mathrm{FAP}<5\,\%$ as uncertain when they were not supported by an independent dataset.

\subsection{Gaussian process modelling}

Stellar activity signals are subject to changing amplitudes, phase shifts, and period shifts around the true rotation period. Fourier-based methods such as the GLS periodogram are not able to model these effects, leading to multiple periodogram peaks at similar periods and residual periodicity after subtracting the strongest signal. This makes finding the true underlying period and assessing its uncertainty difficult. A Gaussian process (GP) is a stochastic model of random variables, described by a mean and a covariance (kernel) function. This kernel function can be used to model the quasi-periodic behaviour of stellar activity. For this purpose, \citet{foreman-mackey2017} proposed a kernel of the form,

\begin{equation}
    k(t_n, t_m) =
\frac{a}{2 + b} \, e^{-c \tau}
\left[
\cos\left( \frac{2\pi \tau}{P} \right)
+ (1 + b)
\right]
+ \delta_{nm}\sigma^2
\end{equation}

where $t_i$ are the individual time points and $\tau = |t_n - t_m|$. It is characterised by an exponential decay on the timescale, $1/c$, describing the evolution and decay of spots, a periodic term with period, $P$, modelling the rotation, and a secular drift. The amplitude is described by $a$, and $b$ models the relative strength of the periodic vs. exponential term. We used the \texttt{celerite} Python implementation of this kernel \citep{foreman-mackey2017}. Following \citet{fuhrmeister2019}, we augmented the kernel by a jitter term, with the parameter $\sigma$ describing additional white noise and added the nightly errors in quadrature.
The advantage of this kernel function is that the covariance matrix can be inverted in $O(n\log(n))$ operations, significantly speeding up the optimisation process. A kernel of this form has been used by \citet{schroder2020} to model stellar rotation in K-giant activity data.
We found the maximum likelihood values of $a$, $b$, $c$, $\sigma$, and $P$ using the SciPy implementation \citep{virtanen2020} of the L-BFGS-B algorithm \citep{byrd1995}. We then used the emcee package \citep{foreman-mackey2013} to assess the posterior probability distribution with Markov chain Monte Carlo (MCMC). We ran 32 walkers for 3000 steps each, 1000 of which were discarded as burn-in.

To account for the inherent flexibility of GPs, we assessed whether a periodic component was justified by comparing the posterior odds of a kernel with a periodic term to an otherwise identical kernel without it. We used the Schwarz criterion, 

\begin{equation}
    S_{\mathrm{GP}} = \max(\log \mathcal{L}_1) - \max(\log \mathcal{L}_2) - \frac{1}{2} d \log n,
\end{equation}

with the likelihood functions, $\mathcal{L}_i$, of the periodic (1) and non-periodic (2) kernels, $d=2$, and $n$ the number of data points. We adopted a threshold of $S_{\mathrm{GP}} \geq 4.6$ in favour of the periodic term for a significant fit, following \citet{fuhrmeister2019}.

\section{Spot modelling}
\label{sec:spotmodel}

For the stars that show modulation at the same period in photometry and RV, we can check whether both are consistent with rotational spot modulation, using the starspot model of the Spot Oscillation And Planet (SOAP) code \citep{boisse2012}. The models are rough consistency checks of the amplitudes, not fits: they assume a single active region at a fixed latitude and contrast, use the solar line profiles, convective blueshift, and plage brightness built into SOAP, and compare datasets taken up to a decade apart, between which the active regions will have changed. The main parameters are the temperature contrast, $\Delta T = T_{\text{star}} - T_{\text{spot}}$, and the spot size, $A$, which gives the filling factor for a single spot. It is the area of the spot as a fraction of the visible hemisphere, $2\pi R_*^2$, and it is thus dimensionless. SOAP specifies a spot by its radius, $r$, in units of $R_*$, so that $A = r^2/2$. Each model was inclined to the median of the inclinations $\sin i = P_\mathrm{rot}/P_\mathrm{max}$, obtained by drawing $P_\mathrm{rot}$ from a Gaussian and $P_\mathrm{max}$ from a split-Gaussian with asymmetric $1\sigma$ widths, keeping only the draws with $P_\mathrm{rot}<P_\mathrm{max}$, since the model assumes that the measured period is the rotation period. The spot was placed at a latitude of $30^\circ$, typical of solar active regions. The spot was sized so that the standard deviation of the model over one rotation equalled that of the best-fitting sine, $K/\sqrt{2}$, in the dataset in which the periodic signal was found, and the model then predicted the signal in the other dataset. Each model was also computed at the 16th and 84th percentiles of these inclinations. As seen in \citet{berdyugina2005}, the temperature contrast rises with photospheric temperature, with active G--K giants having a typical $\Delta T$ around $1000$\,K and typical filling factors between 20--40\,\%, mostly measured as the maximum fraction of the projected stellar disc, which for a single spot is up to twice its fraction $A$ of the visible hemisphere. These measurements are however based on rapidly rotating RS CVn binaries and FK Com giants. For more slowly rotating, single giants, spot properties have never been measured, so it is questionable whether such large spot-filling factors would also apply. 

We adopted contrasts of 600--900\,K (Table~\ref{tab:spots}). We tested values of $\Delta T$ between 500 and 1200\,K, resulting in a change in the ratios of predicted to observed scatter by less than 5\,\%, while the spot sizes change by up to a factor of two. We also modelled an active region in which the spot is surrounded by a plage, which SOAP treats as a bright region with suppressed convective blueshift \citep{dumusque2014}. A plage shifts the RVs while barely changing the brightness and partly cancels the dimming by the spot; the solar convective blueshift used by SOAP is close to that of most evolved stars \citep{liebing2023}. The plage was centred on the spot, with the smallest plage-to-spot area ratio that brought the predicted amplitude to the observed one (Table~\ref{tab:spots}). The individual results of the spot modelling are summarised in Table~\ref{tab:spots} and plotted in Appendix~\ref{app:spotmodels} for three representative stars and for all stars in the supplementary material.

\section{Rotation periods}
\label{sec:rotation_periods}

The detrended S index data for all stars with at least 10 nights were searched for periodic behaviour using GLS periodograms. Whenever a significant period was found, the best fitting sine at this period was subtracted. The periodogram of the residuals was again searched for the most significant peak. This procedure was repeated until there were no significant peaks left. We ran a GP fit with a bounded period prior, uniform in $\log P$, placed around the strongest GLS period, which is shown when the period posterior distribution shows a clear peak. The GP period was adopted when the fit was significant ($S_{\mathrm{GP}} \geq 4.6$). In two cases ($\gamma$ Hya and HIP 96459) with $S_{\mathrm{GP}} < 4.6$, the GP period was adopted after visual inspection, when it agreed with a GLS peak in the S index. In these cases the GP period was preferred over the GLS peak, as the quasi-periodic kernel accounts for the finite lifetime of active regions, which a strictly periodic GLS fit cannot capture. The \textsc{Hipparcos} photometry and the Lick RV data were analysed in the same way. After visual inspection of the periodograms and phase-folded data, 31 stars with significant peaks in at least one of the datasets remained. The significant periods were then compared with the conservative upper limit on the rotation periods derived from $v \sin i$ and $R_*$ (see Sect.~\ref{sec:sample}) to identify possible rotation periods. Four adopted periods lie above this conservative limit, but not by much: taking the uncertainties of both the period and $P_\mathrm{max}$ into account, those of $\eta$ Cep, 52 Cyg, and HIP 96459 are about $1.5\,\sigma$ above the nominal $P_\mathrm{max}$, and that of $\upsilon$ Gem is $2.3\,\sigma$ above it. These are discussed individually in Sect.~\ref{sec:individual_stars} and Appendix~\ref{app:individual_stars}.

The results are summarised in Table~\ref{tab:results}. The periodograms and GP fits for all stars can be found in the supplementary material. Figure~\ref{fig:p_rot_hist} shows the distribution of all plausible rotation periods together with their conservative upper limits. 

\begin{figure}[ht]
    \centering
    \includegraphics[width=.95\linewidth]{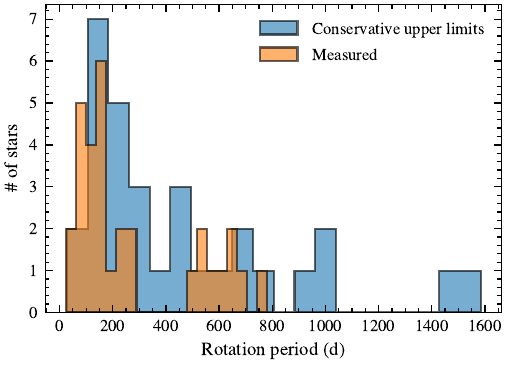}
    \caption{Distribution of the measured rotation periods (orange histogram). The blue histogram shows the conservative upper limits on the rotation periods of the same stars, $P_\mathrm{max}+1\sigma$ (Sect.~\ref{sec:sample}).}
    \label{fig:p_rot_hist}
\end{figure}

We found 11 stars that have RV periods concurrent with a similar S index period. These are HIP 3419, HIP 16335, HIP 21421, HIP 23015, HIP 31700, HIP 36962, HIP 46390, HIP 77512, HIP 96459, HIP 102422, and HIP 107315.

In nine cases (HIP 1562, HIP 3419, HIP 20252, HIP 48356, HIP 77512, HIP 87933, HIP 100064, HIP 105515, and HIP 106481) we found similar periods in S index and \textsc{Hipparcos} photometry. For three further stars (HIP 4906, HIP 60646, and HIP 117503), we found photometric periods that would be consistent with a rotation period. For HIP 4906 ($\varepsilon$ Psc), the photometric period is half of the RV period published by \citet{teng2022}, which we adopted as the rotation period (Sect.~\ref{sec:eps_psc}).

For ten stars, S index periods were found that could be possible rotation periods, but lack additional evidence from other sources. Across the full sample, five stars (77 Tau, $\nu^3$ CMa, $\delta$ CrB, $\iota$ Cap, and OU And) confirm or refine previously published periods. While the absence of a photometric or RV counterpart limits the confidence in these periods, their consistency with the $v \sin i$-based upper limits makes a rotational origin plausible. The individual results for some of these stars are discussed in Appendix~\ref{app:individual_stars}.
We flag a rotation period as uncertain (U in Table~\ref{tab:results}) if at least one of the following applies: (i) it is found in only one dataset, and its GLS peak there has a permutation FAP above 0.1\,\%; (ii) it exceeds the conservative upper limit, $P_\mathrm{max}+1\sigma$; (iii) the adopted period differs by more than 20\,\% from the period it is matched with in another dataset, also after allowing for a factor of two. This applies to 12 of the 31 periods: $\varepsilon$ Vir, $\gamma$ Hya, $\eta$ Dra, $\theta$ Her, $\alpha$ Sct, and 39 Cyg (i); $\upsilon$ Gem, $\eta$ Cep, and 52 Cyg (ii); HIP 96459 (ii and iii); and 54 Per and $\alpha$ Hya (iii). Even where the GP posterior peaks at the best-fit period, $S_{\mathrm{GP}}$ is mostly below the threshold, because for sparsely observed stars the non-periodic kernel fits the data nearly as well as the periodic one.

\begin{figure}[htpb]
    \centering
    \includegraphics[width=.95\linewidth]{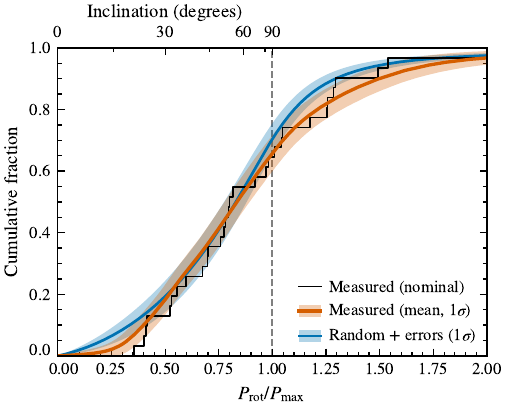}
    \caption{Cumulative distribution of $P_\mathrm{rot}/P_\mathrm{max}$ for the 31 stars with a rotation period. The black curve shows the nominal values. For the orange curve and band, we drew $P_\mathrm{rot}$, $R_*$, and $v\sin i$ of each star $10^4$ times from their uncertainties (Sect.~\ref{sec:sample}) and took the mean and $1\sigma$ scatter of the resulting curves. The blue curve and band show the mean and $1\sigma$ scatter of 4000 simulated samples with isotropic spin axes that went through the same Monte Carlo (see Sect.~\ref{sec:rotation_periods}). The top axis gives the inclination, $i=\arcsin(P_\mathrm{rot}/P_\mathrm{max})$, and the dashed line marks $P_\mathrm{rot}=P_\mathrm{max}$. One star, $\eta$ Cep (nominal $P_\mathrm{rot}/P_\mathrm{max}=2.8$), lies outside the plotted range.}
    \label{fig:inclination_cdf_mc}
\end{figure}

From the measured period and the maximum rotational period, we can infer the inclination angles of the stellar rotation axis, assuming that the measured period is the true equatorial rotation period, as $\sin i = P_\mathrm{rot}/P_\mathrm{max}$. Figure~\ref{fig:inclination_cdf_mc} shows the cumulative distribution of this ratio, with the corresponding inclination on the top axis. Eleven of the 31 stars have a measured period above the nominal $P_\mathrm{max}$, which is largely a result of the large $v \sin i$ uncertainties (median 22\,\%). To compare the distribution with what would be expected from randomly oriented spin axes, we simulated 4000 samples of our 31 stars. In each sample, every star has a random inclination with $\cos i$ uniformly distributed, which together with its measured $P_\mathrm{rot}$ and $R_*$ sets its $v \sin i$. We perturbed these values once with the uncertainties of the star, so that each simulated sample resembles a measured catalogue, and then applied the same Monte Carlo to it as to the measured values, with $10^4$ draws per star. The Monte Carlo curve is consistent with the isotropic expectation: 34\,\% of the draws lie above $P_\mathrm{max}$, compared with $(30\pm5)\,\%$ expected. It lies slightly below the expectation at both ends. Small ratios are rarer than expected from a randomly oriented sample, as rotational modulation is harder or even impossible to detect at low inclinations, and more stars lie above $P_\mathrm{max}$, in particular at $P_\mathrm{rot}/P_\mathrm{max}\approx1.1$--$1.5$. The latter may indicate that some $v \sin i$ values or their uncertainties are inaccurate for these slow rotators or that some periods are harmonics, are not caused by rotation, or do not trace the equatorial rotation. Furthermore, the observing cadence limits the detection of very short and long periods in some of the datasets. Within the uncertainties, we still interpret our inferred inclinations as consistent with a randomly oriented sample. 

\begin{figure}[htpb]
    \centering
    \includegraphics[width=\linewidth]{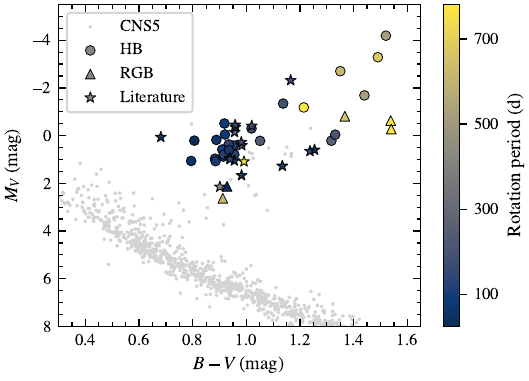}
    \caption{CMD based on \textsc{Hipparcos} photometry, showing all stars for which we found possible rotation periods, with the colour corresponding to their period. G and K giants with literature periods from \citet{choi1995}, \citet{auriere2015}, and \citet{schroder2020} are plotted as star markers. As a reference, we plot the stars of the fifth Catalogue of Nearby Stars \citep{golovin2023}, labelled CNS5, that are also in the \textsc{Hipparcos} catalogue.}
    \label{fig:cmd}
\end{figure}

In Fig.~\ref{fig:cmd}, we show the detected periods on a CMD. It can be seen that the longest ($\gtrsim500$\,d) periods are mostly found at $B-V>1.2$; the only exception is $\eta$ Cep ($B-V=0.91$), whose period is uncertain.

\begin{figure}[htpb]
    \centering
    \includegraphics[width=\linewidth]{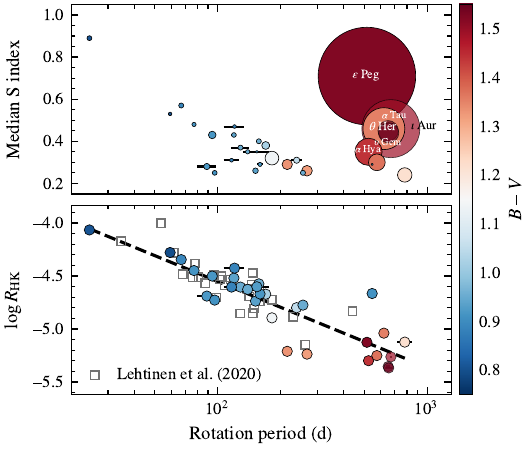}
    \caption{Median S index and $\log R_\mathrm{HK}$ plotted against the rotation period, with the colour giving $B-V$. Horizontal bars show the period uncertainties. \textit{Top:} Median S index. Marker sizes are proportional to stellar radius and stars with $R_*>40\,R_\odot$ are marked with their names. \textit{Bottom:} $\log R_\mathrm{HK}$. All markers have the same size. The dashed line is a linear fit to our stars in logarithmic space ($\log R_\mathrm{HK} = (-0.81\pm0.08) \log P_\mathrm{rot} - (2.93\pm0.19)$, with $P_\mathrm{rot}$ in days).}
    \label{fig:p_rot_s_index}
\end{figure}

\citet{noyes1984} found a strong correlation between chromospheric activity and rotation period in main-sequence stars, which has also been observed in giant stars \citep{auriere2015}. In Fig.~\ref{fig:p_rot_s_index}, we plot the median S index against our measured rotation periods, finding a similar relation for the stars with $R_*<40\,R_\odot$. The largest giants seem to fall out of this relation. The S index is, however, not directly comparable between stars of different colour: towards cooler stars its continuum bands V and R fade relative to the H and K lines, so that a red giant has a higher S index than a yellow one at the same chromospheric flux. We therefore also converted the S index into $R_\mathrm{HK}=1.34\times10^{-4}\,C_\mathrm{cf}\,S$ \citep{noyes1984}, which gives the chromospheric H and K flux relative to the bolometric flux, using the conversion factor for giants, $\log C_\mathrm{cf} = 0.45-0.49\,(B-V)-0.25\,(B-V)^2-0.066\,(B-V)^3$ \citep{rutten1984}. As no photospheric or basal correction is established for giants on this scale, we did not convert further to $R'_\mathrm{HK}$. \citet{lehtinen2020} use the same conversion, but subtract the photospheric flux of main-sequence stars \citep{noyes1984}, which lowers $\log R_\mathrm{HK}$ of our stars by at most 0.08\,dex. In $R_\mathrm{HK}$, all stars follow a single relation ($\rho=-0.86$, $p=8\times10^{-10}$; bottom panel of Fig.~\ref{fig:p_rot_s_index}); the only clear outlier is $\eta$ Cep, whose period is uncertain. The 28 giants of \citet{lehtinen2020} not in our sample follow the same relation. Because the longest periods are found in the reddest stars ($\rho=0.84$ between $P_\mathrm{rot}$ and $B-V$) and $C_\mathrm{cf}$ itself depends on colour, part of this correlation reflects colour; with $B-V$ held fixed, a partial rank correlation of $r=-0.45$ ($p=0.01$) remains. Some of the stars around 100\,d and 200\,d lying below the general trend may again indicate that the measured periods are half their rotation period.

For 16 of the stars with a period in our sample, \citet{lehtinen2020} also measured one from the same HKP data. For 11 of them, the periods agree with ours within $2\sigma$ (Table~\ref{tab:results}). For $\delta$ CrB and $\xi$ Her, their periods of 52 and 106\,d lie at or just below the lower end of the periods we find in individual observing seasons; the differences for $\gamma$ Hya, HIP 96459, and $\theta$ Her are discussed in Appendix~\ref{app:individual_stars}. Their longest period is 440\,d, so none of our periods above 500\,d is in their sample.

\begin{figure}[htpb]
    \centering
    \includegraphics[width=\linewidth]{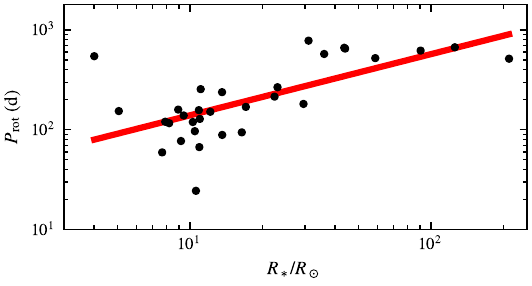}
    \caption{Measured rotation periods plotted against the stellar radius in logarithmic scale. The line is a linear fit in logarithmic space ($\log P_\mathrm{rot} = (0.61\pm0.13) \log R_* + (1.53 \pm 0.17)$).}
    \label{fig:p_rot_r_star}
\end{figure}

The trend of rotation periods increasing with stellar radius can also be observed in Fig.~\ref{fig:p_rot_r_star}. Since $P_\mathrm{max}\propto R_*$ and we required the periods to be compatible with $P_\mathrm{max}$, part of this correlation is expected by construction.

\section{Individual stars}
\label{sec:individual_stars}

In this section, we discuss some of the stars that show possible rotational signals in at least two of the analysed datasets. Some of these were also the subject of previous publications and possible planet host candidates. Further stars are discussed in Appendix~\ref{app:individual_stars}.

\begin{figure}[ht]
    \centering
    \includegraphics[width=\linewidth]{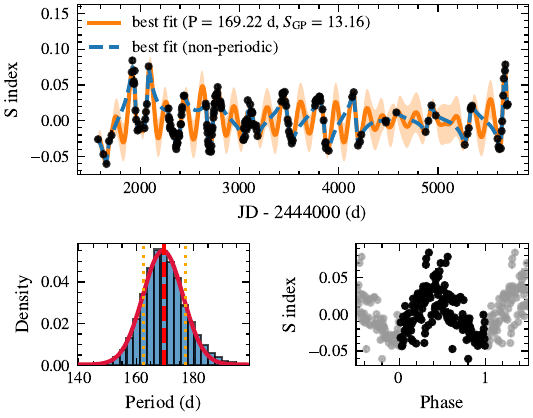}
    \caption{GP modelling of $\beta$ Cet. \textit{Top:} Data (black) with best-fit periodic GP with $1\,\sigma$ credible interval (orange) and best-fit non-periodic GP (blue dashed). \textit{Bottom left:} MCMC posterior distribution of rotation period with Gaussian fit and $1\sigma$ credible bounds. \textit{Bottom right:} Phase-folded light curve at the best-fit period. The nightly S index was high-pass filtered with $P_{\rm cut}=1000$\,d before the fit.}
    \label{fig:gp_beta_cet_main}
\end{figure}

\subsection{\texorpdfstring{$\beta$ Cet}{beta Cet} (HIP 3419 / HD 4128)}
\label{sec:beta_cet}

The periodogram of the S index data shows a strong peak at $(170.91\pm0.28)$\,d and the GP period is $(169.6^{+7.8}_{-6.9})$\,d, with $S_{\mathrm{GP}}=13.2$ (Fig.~\ref{fig:gp_beta_cet_main}). After prewhitening, a weaker peak remains at $(83.16\pm0.15)$\,d, half the rotation period. The periodogram of the \textsc{Hipparcos} photometry contains a significant peak at $(169.4\pm4.5)$\,d, in good agreement with the S index period.

\begin{figure}[ht]
    \centering
    \includegraphics[width=1\linewidth]{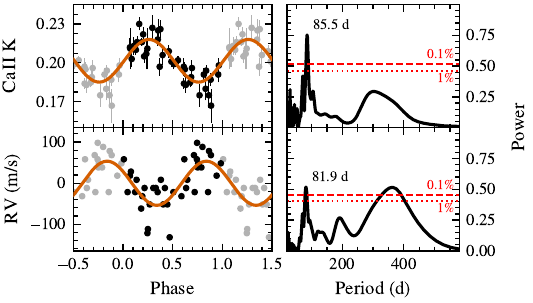}
    \caption{Periodic analysis of the data from \citet{tsvetkova2013}. \textit{Top:} Ca\,\textsc{ii} K line strength relative to the continuum intensity at 395\,nm. \textit{Bottom:} Radial velocity, relative to its mean. The left panels show the data with the best-fitting sine, phase-folded and duplicated to cover a phase from $-0.5$ to 1.5. The right panels show the GLS periodograms, whose highest peak gives the period; the dashed and dotted lines indicate permutation FAPs of 0.1 and 1\,\%.}
    \label{fig:betacet_tsvek}
\end{figure}

\citet{tsvetkova2013} inferred a significantly larger period of 215\,d from fitting a magnetic model to the variation of the Stokes V profile. The maximum period from $v \sin i$ is ($210^{+30}_{-24}$)\,d, placing both periods in a possible range. \citet{massarotti2008} found a slightly larger $v \sin i$ of 5.8\,km\,s$^{-1}$, in favour of the shorter period. \citet{tsvetkova2013} included measurements of the Ca\,\textsc{ii} K flux and RV in their paper that were extracted from the spectra they used to determine the rotation period. The periodograms of these time series both show periods just above 80\,d, or about half of the S index and photometric period (see Fig.~\ref{fig:betacet_tsvek}), and a peak at this half period is also present in the HKP S index. Since two sets of spots at opposite longitudes can produce a signal at half the rotation period, the 170\,d period is more likely to be the true rotation period.

The strongest period in the Lick RVs is ($181.1\pm1.1$)\,d, which is close to the 170\,d period, but also close to a half year. The periodogram shows a peak at 361\,d with a similar power, which also appears in the window function. When fitting and subtracting the 361\,d period, the other peak also vanishes, suggesting that both might be related to the observing cadence. The folded RV data do, however, show nearly sinusoidal modulation with an amplitude around 100\,m\,s$^{-1}$, similar to the RVs from \citet{tsvetkova2013}. A single spot ($\Delta T = 800$\,K) sized to the photometry ($A=0.77$\,\%) predicts about three quarters of the RV scatter (Table~\ref{tab:spots}, Fig.~\ref{fig:betacet_soapmodel}). If the spot is surrounded by a plage with 3.6 times its area, the model reproduces both the photometric and the RV amplitude.
In the case that the RV signal is real and caused by rotation, the 10\,d difference to the other periods could be explained by the migration of spots and differential rotation, as there is a gap of about 5.5 years between the HKP and Lick datasets.

\subsection{\texorpdfstring{$\alpha$ Tau}{alpha Tau} (HIP 21421 / HD 29139)}
\label{sec:aldebaran}

\begin{figure}[!htpb]
    \centering
    \includegraphics[width=1\linewidth]{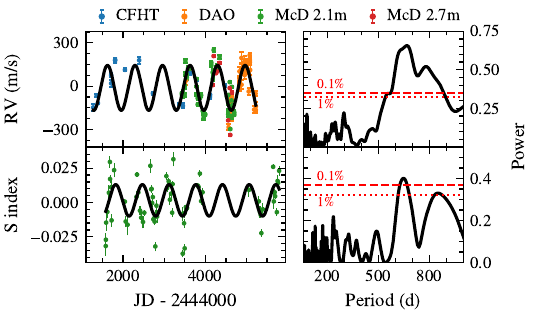}
    \caption{Overlapping RV (top) and S index (bottom) data for $\alpha$ Tau, with the best-fitting sine (left), and their GLS periodograms (right). The RV sources are indicated in the legend and described in \citet{hatzes2015}. The S index is given as nightly means and was high-pass filtered with $P_{\rm cut}=2000$\,d. The dashed and dotted lines in the periodograms indicate permutation FAPs of 0.1 and 1\,\%.}
    \label{fig:Aldebaran_RV_S_periodograms}
\end{figure}

Aldebaran ($\alpha$ Tau) has been the subject of discussion surrounding a possible planetary companion. \citet{hatzes2015} found RV variations and attributed these to a planet with an orbital period of ($628.96\pm0.90$)\,d. Additionally, they suggested that a $\approx 520$\,d period in the residual RVs and the activity indicators Ca\,\textsc{ii} K, S index, and H$\alpha$ could be caused by the rotation of the star. Although their favoured explanation for the long-term RV variation was a planetary companion, they also discussed the possibility that both periods come from rotational modulation. The deviation between the periods can be the result of differential rotation and the migration of stellar surface features. \citet{reichert2019} used the Lick RV measurements and combined them with the dataset from \citet{hatzes2015} to refute the planet hypothesis for $\alpha$ Tau. They argued against the existence of this planet, noting that the periodogram power for this period is not stable over a longer time. The S index data, ranging from September 1983 to March 1995, were taken long before the Lick RVs, but coincide with part of the combined RV dataset from \citet{hatzes2015}. The overlapping RV data were taken between November 1982 and August 1993. The strongest periodogram peaks for this part of the RV data and the S index data are ($668.4\pm6.3$)\,d and ($648\pm12$)\,d, respectively (see Fig.~\ref{fig:Aldebaran_RV_S_periodograms}). The Lick RVs, taken about a decade later, have their strongest peak at $(591\pm5)$\,d, followed by peaks at 797 and 428\,d.

With the $v \sin i$ from \citet{massarotti2008} and an assumed error of 1\,km\,s$^{-1}$, the maximum rotation period is at $517^{+157}_{-97}$\,d. In the case that the RV variations are mainly the result of spot modulation, we would however expect significant photometric variability. A single SOAP spot sized to the semi-amplitude of 133\,m\,s$^{-1}$ of the best-fitting sine to the Lick RVs predicts a photometric scatter six times that of the \textsc{Hipparcos} photometry, which overlaps part of the RV dataset of \citet{hatzes2015}. Surrounding the spot with a plage can lower this at most to five times within the region sizes that SOAP can model (Table~\ref{tab:spots}). Despite this discrepancy, the concurrent RV and S index periods still support intrinsic stellar activity as their origin.

\subsection{\texorpdfstring{$\iota$ Aur}{iota Aur} (HIP 23015 / HD 31398)}
\label{sec:iota_aur}

The bright K\,II giant $\iota$ Aur has a radius of $126\,R_{\odot}$. The RV data show periodic variations with a varying semi-amplitude up to 500\,m\,s$^{-1}$. The highest periodogram power of the Lick RVs is at ($702\pm11$)\,d but the behaviour is clearly multiperiodic, with further significant peaks at 922, 1261, 2097, and 555\,d. A GP fit to the RVs is significant ($S_{\mathrm{GP}}=8.4$), but places the period at $(814^{+75}_{-59})$\,d, between the 702 and 922\,d peaks, since the RV variations change their shape from cycle to cycle.

The S index data show a period of ($670.6\pm7.6$)\,d and the maximum rotation period from $v \sin i$ is ($855^{+148}_{-117}$)\,d, so a period around 700\,d is consistent with this limit. However, strong photometric variability would be expected if these variations were caused by spot modulation.

A single SOAP spot, 900\,K cooler than the photosphere, on a star rotating at the RV period and inclined to the inferred $i=51^\circ$, needs a spot size of $A=3.4$\,\% for its RV scatter to match that of the best-fitting sine, with a semi-amplitude of 192\,m\,s$^{-1}$ (Fig.~\ref{fig:iotaaur_soapmodel}). Such a spot would produce a photometric scatter (standard deviation) five to six times the 0.005\,mag of the \textsc{Hipparcos} photometry over the range of inclinations. Surrounding the spot with a plage does not reduce this within realistic region sizes (Table~\ref{tab:spots}). The \textsc{Hipparcos} observations predate the RVs by about a decade, so a lower activity level at that time is possible, but spots are unlikely to explain the full RV amplitude alone. Considering the quasi-periodic variations in RV and S index, we still see activity as their most likely origin.

\subsection{\texorpdfstring{$\upsilon$ Gem}{ups Gem} (HIP 36962 / HD 60522)}
\label{sec:ups_gem}

The luminous K5\,III giant $\upsilon$ Gem has a radius of about $40\,R_\odot$. The S index shows a significant periodogram peak at $(645\pm11)$\,d (FAP between 1 and 5\,\%). The Lick RVs show a single strong period at $(663\pm10)$\,d with a semi-amplitude of $\approx100$\,m\,s$^{-1}$, close to the dominant S index period and pointing to a common stellar origin. A GP fit to the RVs is significant, with $P=(655^{+25}_{-22})$\,d and $S_{\mathrm{GP}}=8.2$, whereas the S index GP is not ($S_{\mathrm{GP}}=-3.8$). We adopted the RV GP period as the rotation period. The \textsc{Hipparcos} photometry shows a dominant periodogram peak at $(96\pm1)$\,d and broad power around 700\,d.

The upper limit on the rotation period is $425^{+102}_{-70}$\,d. The $\approx650$\,d S index/RV signal exceeds this limit by $2.3\,\sigma$. A rotational origin would therefore require the $v \sin i$ of this star to be overestimated, so we flagged it as uncertain. In radius, period, and activity level, $\upsilon$ Gem closely resembles $\alpha$ Tau (Fig.~\ref{fig:p_rot_s_index}). They are also both on the RGB and at similar positions in the CMD (Fig.~\ref{fig:cmd}). Given this resemblance, it is plausible that the same mechanism underlies the $\approx650$\,d variability in both stars -- either rotational modulation, a somewhat longer activity cycle, or stellar oscillations.

\subsection{\texorpdfstring{$\delta$ CrB}{delta CrB} (HIP 77512 / HD 141714)}
\label{sec:delta_crb}

\begin{figure}[htpb]
    \centering
    \includegraphics[width=\linewidth]{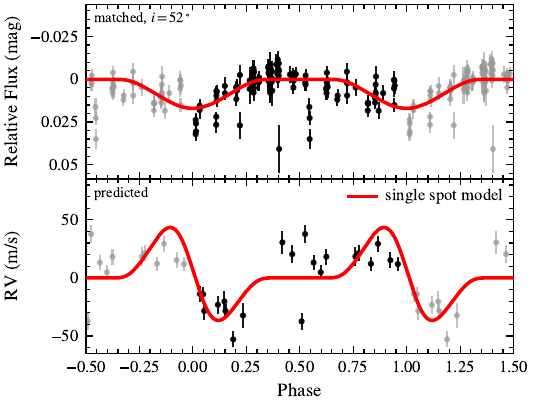}
    \caption{\textsc{Hipparcos} light curve (top) and Lick RVs (bottom) of $\delta$ CrB, phase folded at their respective best-fit periods (grey data points were duplicated for visual purposes). The SOAP model is sized so that its standard deviation over one rotation matches that of the best-fitting sine to the photometry, predicting the RV curve. The red curve shows a single spot of $A=1.3$\,\% at a latitude of $30^\circ$, with $\Delta T = 600$\,K and $i=52^\circ$, phased separately to each dataset. The model is not intended as a fit to the data, but rather to compare the photometric and RV amplitudes expected from spot modulation.}
    \label{fig:delta_crb_soapmodel}
\end{figure}

The periodogram of the S index data for $\delta$ CrB shows a strong peak at $(58.84\pm0.04)$\,d. This period is consistent with the period of $(59\pm2)$\,d from \citet{choi1995} (derived from S index and photometry) and with the photometric period of 57\,d from \citet{strassmeier1999}. All these periods are consistent with rotation, considering the maximum period of $(75^{+9}_{-7})$\,d from $v \sin i$. The GP model of the S index data gives a period of $(59.3^{+0.6}_{-0.5})$\,d, with $S_{\mathrm{GP}}=71$. Fitted season by season, the S index gives periods between 51 and 78\,d, scattered around this value.
The \textsc{Hipparcos} photometry of $\delta$ CrB also shows significant periodicity with the highest-power periodogram peak at $(57.22\pm0.25$)\,d, close to the S index period. The difference can be attributed to a slightly shifting period through differential rotation, as these data only span a small part of the HKP observation period. The photometric observations of \citet{choi1995} show a larger amplitude than the \textsc{Hipparcos} observations, which is consistent with the S index being higher and more variable during their observation period, compared to the \textsc{Hipparcos} observation period.
The Lick RV data for this star are sparse, with only 21 data points taken over 7 years. The highest peak is at a period of $(59.91\pm0.22)$\,d. However, it does not reach the 5\,\% permutation FAP level (analytic FAP 3.2\,\%). While this peak on its own should not be trusted, the fact that it is close to the other periods supports that some RV modulation could have been produced by rotation. A SOAP spot model matching the photometric amplitude of the \textsc{Hipparcos} data predicts an RV amplitude that roughly matches the Lick data (Fig.~\ref{fig:delta_crb_soapmodel}), supporting the spot modulation hypothesis. Since the RVs were taken several years after the \textsc{Hipparcos} photometry, a slightly different period and amplitude can be expected.

\subsection{\texorpdfstring{$\xi$ Her}{xi Her} (HIP 87933 / HD 163993)}
\label{sec:xi_her}

The star $\xi$ Her is classified as a periodic variable in the \textsc{Hipparcos} catalogue. A period of $(121.1\pm1.0)$\,d is inferred from the periodogram of the \textsc{Hipparcos} photometry with an amplitude of $\approx 0.03$\,mag, together with its first harmonic at $(60.3\pm0.3)$\,d. The same period is visible in the S index data. The periodogram analysis shows a strong peak at $(120.1\pm0.2)$\,d and weaker residual peaks at 85 and 97\,d. The GP modelling leads to a final S index period of $(119.6\pm3.3)$\,d, with $S_{\mathrm{GP}}=25.9$, in agreement with the photometric period. In the four observing seasons in which the period is resolved, the S index gives periods between 108 and 127\,d. The upper limit on the rotation period is $P_\mathrm{max} = (115^{+17}_{-13})$\,d, which further supports that the 120\,d period can be explained by rotational modulation.

The \textsc{Hipparcos} observations roughly coincide with an overall increase of the S index (i.e. increased stellar activity), which could explain the relatively large photometric amplitude. Using SOAP, a single spot of size $A=1.8$\,\% and $\Delta T = 600$\,K can recreate the photometric amplitude of about 0.03\,mag, while also matching the shape of the folded light curve (see the supplementary material). This model leads to RV variability with a standard deviation of 20\,m\,s$^{-1}$. The Lick RVs scatter by 25\,m\,s$^{-1}$, and adding a plage to the model would only increase its RV signal, so it is plausible that the rotational signal is hidden in the scatter of these data, especially considering that the star's activity may have decreased during the Lick observation period compared to the high activity levels during the \textsc{Hipparcos} observations. The average spacing of the Lick observations is about 90\,d, which means that the period would not be well resolved, if it were present.

There is another period that appears in the S index, at $(1906\pm99)$\,d, when the undetrended nightly data were searched after subtracting the long cycle. It is also visible in the RVs at $(1917\pm130)$\,d, with a FAP between 1 and 5\,\%. This period could indicate another stellar activity cycle on top of the longer $(5400\pm420)$\,d S index cycle.

\subsection{\texorpdfstring{$\varepsilon$ Peg}{epsilon Peg} (HIP 107315 / HD 206778)}
\label{sec:eps_peg}

The supergiant $\varepsilon$ Peg (K2\,Ib) has a radius of about $211\,R_{\odot}$. It shows a high S index, varying between 0.6 and 0.8, although its chromospheric flux relative to the bolometric flux ($\log R_\mathrm{HK}=-5.1$) is at the lower end of our sample, and an RV semi-amplitude up to $1$\,km\,s$^{-1}$. The highest-power periods in the S index and RV data are ($585\pm7$)\,d and ($515\pm7$)\,d, respectively. Another period around 117\,d appears in \textsc{Hipparcos} photometry, and twice this period, 234\,d in the RVs.

The matching RV and S index periods, together with the variability of the S index, point to activity as their common origin. The maximum rotation period is $(1281^{+303}_{-208})$\,d. For the $\approx 500$\,d period to be caused by rotational modulation, either the rotation axis would have to be inclined by $\approx 25^{\circ}$, or the measured period would be half the true rotation period, or the $v \sin i$ measurement would be inaccurate. In the first case, a single SOAP spot, 900\,K cooler than the photosphere and seen at the inferred $i=24^\circ$, needs a spot size of $A=5.1$\,\% for its RV scatter to match that of the best-fitting sine, with a semi-amplitude of 324\,m\,s$^{-1}$ (supplementary material). It predicts a photometric scatter (standard deviation) 2.4 times the 0.012\,mag of the \textsc{Hipparcos} photometry, and about as much when the spot is surrounded by a plage (Table~\ref{tab:spots}). In the second case, two spots at opposite longitudes on a star rotating in 1031\,d, seen at the inferred $i=51^\circ$, need $A=3.6$\,\% each and predict 1.6 times the photometric scatter. Since there is no temporal overlap with the RVs, a lower activity level at the time of the \textsc{Hipparcos} observations could explain this difference, but the RV variations of up to 1\,km\,s$^{-1}$ in the most active part of the time series would require correspondingly larger spots.

\section{Discussion}
\label{sec:discussion}

Measuring rotational starspot modulation in giant stars is possible, but it requires precise photometric data with a long time baseline for bright stars, which is hard to find. With most surveys targeting fainter stars or observing only in short segments, the \textsc{Hipparcos} photometry still seems to be one of the best sources for bright stars to this day, even though it is more than 30 years old. Another future source for photometric data is the epoch photometry that will be released with Gaia Data Release 4 (DR4) by the end of 2026\footnote{\href{https://www.cosmos.esa.int/web/gaia/release}{https://www.cosmos.esa.int/web/gaia/release}}.

All adopted periods are compatible with the upper limits on the rotation period derived from $v \sin i$ and the stellar radii (Sect.~\ref{sec:rotation_periods}), making them plausible rotation periods. However, because the HKP S index, \textsc{Hipparcos} photometry, and Lick RV data each probe different physical layers of a star, we discuss here the different lines of evidence for and against a rotational origin of the detected signals.

\subsection{Evidence supporting a rotational origin}

When the same (or similar) period is found in independent datasets, the signals are likely caused by a common stellar origin. While we see rotation as a possible explanation for many of the discussed stars, we cannot, however, rule out other stellar phenomena like pulsations in some cases, based on the available data alone.

Transient phenomena such as individual spot emergence or decay events can produce quasi-periodic signals on short timescales, but these are generally not stable over many rotation cycles. In several of the well-sampled stars (e.g. $\xi$ Her, $\iota$ Cap, and $\delta$ CrB), the period remains consistent across observing seasons spanning many years, pointing to a stable underlying phenomenon.

Figure~\ref{fig:p_rot_s_index} shows that stars with shorter measured periods tend to have higher chromospheric activity. This activity--rotation relationship is a well-established feature of main-sequence stars \citep{suarezmascareno2015} and has been documented for giant stars as well \citep{auriere2015, dixon2020}. The fact that our sample of G and K giants follows the same trend lends further credibility to the rotational interpretation of the detected periods.

For the two stars for which the same period is found in both \textsc{Hipparcos} photometry and RV ($\delta$ CrB and $\beta$ Cet), the photometric and RV amplitudes are consistent with a single-spot configuration of small filling factor and moderate temperature contrast. For $\delta$ CrB, a single spot sized to the photometry predicts about the observed RV scatter. For $\beta$ Cet, it predicts three quarters of the RV scatter, and a plage of a few times the spot's area, as on the Sun, is enough to reproduce both amplitudes. This provides a self-consistent physical picture: the same active region that produces the photometric variability generates the observed RV amplitude.

While strictly periodic signals are expected for planetary companions, stellar surface activity produces quasi-periodic signals whose phase, amplitude, and period shift gradually as active regions evolve and migrate. This is seen in many of the observed stars. The gradual period drifts visible in some stars' periodograms are consistent with differential rotation causing spots at different latitudes to show slightly different apparent periods \citep{baliunas1985, donahue1996}.

\subsection{Alternative explanations and caveats}

Despite these arguments, several considerations call for caution when interpreting the detected periods as rotation periods. Single giants, as opposed to rapidly rotating RS CVn binaries, typically exhibit low activity levels and small photometric variability. The detection of rotational modulation in such low-amplitude signals is inherently difficult, and the risk of finding spurious periods in the noise is higher than for active main-sequence stars. The large fraction of uncertain or marginally significant periods reflects this difficulty.

As demonstrated for many Sun-like stars by the Mount Wilson survey, solar-type activity cycles with periods of several years to decades are common, modulating the overall level of chromospheric emission \citep{baliunas1995a, baliunas1998}. The data used here reveal long-term trends in the S index of many sample stars, which we attempted to remove by subtracting a high-pass cosine series (Sect.~\ref{sec:obs}). Some of the measured periods could also be the result of higher harmonics of these long activity cycles. Furthermore, spots distributed at several active longitudes can produce a signal at the first or higher harmonics of the rotation period.

Surface spots are not static: they emerge, grow to a maximum area, and then decay. For solar-type stars, individual spot lifetimes range from days (small spots) to months or even years for large active complexes \citep{hussain2002, giles2017}. On giant stars with their much larger convective cells, spots or active regions may live longer, but the individual lifetimes and migration rates are poorly known. When a dominant spot emerges, lives for a few rotation periods, and then decays, the resulting quasi-periodic signal may closely resemble rotational modulation, but what we measure would be related to the lifetime of the active region rather than to stellar rotation. We note that p-mode (solar-like) oscillations are present in all giant stars and produce RV jitter of order 10--20\,m\,s$^{-1}$ for relatively stable K giants, increasing strongly with luminosity \citep{hekker2008}. Their periods of hours to a few days \citep{frandsen2002, kjeldsen1995} are much shorter than those we detect, so they add scatter around a long-period signal but cannot produce it.

A specific non-radial mechanism, oscillatory convective modes (highly non-adiabatic $g^-$ modes, i.e. the g-mode branch of convectively unstable layers), has been proposed to explain the long secondary periods seen in luminous red giant variables \citep{saio2015}. These modes are expected in stars with $\log(L/L_\odot) \gtrsim 3$ and produce periods of several hundred to over 1500\,d, increasing with luminosity. Two of our stars, $\iota$ Aur and $\varepsilon$ Peg, lie well above this luminosity threshold, with $\log(L/L_\odot) \approx 3.6$ and $3.9$, respectively, and must therefore be considered candidates for oscillatory convective mode activity. A characteristic signature of multi-mode pulsation is the presence of several incommensurable periods \citep{hatzes1999, spaeth2024}, which is indeed what the two supergiants show. The absence of a simple harmonic structure in their periodograms may therefore point to pulsation as an alternative or additional explanation for their RV variability. The less luminous stars in our sample are not the main candidates for oscillatory convective modes, strengthening the rotational interpretation there, though non-radial pulsations in general cannot be excluded yet. The spot models point the same way: for the five stars with RV semi-amplitudes above 100\,m\,s$^{-1}$ and periods above 400\,d ($\sigma$ Per, $\alpha$ Tau, $\iota$ Aur, $\upsilon$ Gem, and $\varepsilon$ Peg), a spot sized to the RVs predicts several times the observed photometric scatter, and a plage could close this gap only with a coverage that is implausible for their low chromospheric flux (Table~\ref{tab:spots}). These models are, however, only rough estimates (Sect.~\ref{sec:spotmodel}).

While the inclination distribution as a whole is consistent with random orientations (Sect.~\ref{sec:rotation_periods}), several stars imply $20$--$25^{\circ}$ (e.g. 54 Per, $\gamma$ Hya, and $\varepsilon$ Peg), at which little rotational modulation is expected. A significant detection therefore argues that the measured period is not the rotation period or is a harmonic of it, or that the $v \sin i$ uncertainty allows a larger inclination. Largely deviating $v \sin i$ measurements in the literature for the same star further reflect the difficulty of measuring rotational velocities in slow rotators.

\section{Summary}
\label{sec:summary}

The goal of this study was to find rotation periods for G and K giants from the Lick sample using archival \textsc{Hipparcos} epoch photometry, Mount Wilson S index data, and the Lick RV measurements, and to assess whether rotational modulation could account for some of the observed RV variations.

We searched for periodic signals using GLS periodograms and GP modelling with a quasi-periodic kernel, validating results between data sources and against upper limits on the rotation period derived from $v \sin i$ and stellar radii. Plausible rotation periods were found for a total of 31 stars. The strongest evidence comes from nine stars ($\delta$ CrB, $\iota$ Cap, $\xi$ Her, $\beta$ Cet, $\upsilon^1$ Hya, $\alpha^2$ Cap, $\iota$ Cet, 54 Per, and $\rho$ Cyg) for which consistent periods were recovered from both the S index and \textsc{Hipparcos} photometry. For $\delta$ CrB and $\beta$ Cet, a compatible period also appears in the Lick RV data (for $\delta$ CrB below the 5\,\% FAP level), and simple SOAP spot models show that the photometric and RV amplitudes can be explained by a rotating spot configuration. For a further nine stars, the same period appears in the S index and the Lick RVs, supporting a rotational origin. In some of these cases, the RV amplitude requires relatively high spot-filling factors to be explainable by spot modulation alone, which would lead to higher than observed photometric variations. Since real spot sizes are not readily known for single K giants, additional mechanisms such as oscillatory convective modes should still be investigated. Another ten stars show a period only in the S index that is consistent with the $v \sin i$-based upper limit, and three show a period only in photometry, for $\varepsilon$ Psc at half of a published RV period. For 12 stars, the identified rotation period remains uncertain, because it rests on a single marginal detection, exceeds the $v \sin i$-based limit, or differs between datasets. 

The measured rotation periods range from about 25\,d (OU And) to roughly 800\,d ($\iota$ Cet). Most of the periods (20 of 31) are $\lesssim250$\,d, in agreement with previous studies that found rotational modulation to be detectable mainly in giants with moderate to fast rotation rates. Stars with shorter periods tend to show higher chromospheric activity, consistent with the known activity--rotation relationship. In the S index, the largest giants appear as exceptions, but this is a consequence of their red colours, and in $R_\mathrm{HK}$, they follow the same relation. The distribution of $P_\mathrm{rot}/P_\mathrm{max}$, from which the inclination angles follow, is consistent with random orientations once the $v \sin i$ uncertainties are taken into account, with a hint of more stars than expected having periods above $P_\mathrm{max}$. Several stars also show long-term S index trends consistent with magnetic activity cycles on timescales of roughly $4$--$15$\,yr, although an even longer dataset would be needed to confirm their periodicity.

Further findings include that the supergiants and bright giants show more variability, especially in RV, complicating the search for companions around these stars. The largest giants in our sample show complicated RV variations with high, variable amplitudes and a variable S index.

The results presented here can serve as a guide when evaluating planetary candidates from this sample: any candidate period coinciding with a rotation period identified here should be subject to additional scrutiny before a substellar companion is claimed. Forthcoming Gaia DR4 epoch photometry will provide a valuable independent check for many of these targets. Furthermore, photometry with contemporaneous spectroscopy, line-profile indicators, and spot models adapted to giants could be used to investigate the active regions of these stars in more detail.

\section*{Data availability}
The supplementary material, which contains the period search figures of all stars discussed in this paper (periodograms, detrending, and GP fits) and the SOAP spot models, is available on Zenodo at \url{https://doi.org/10.5281/zenodo.24287024}.

\begin{acknowledgements}

We thank the referee, Robert Wittenmyer, for helpful comments that improved this paper. SJ thanks Delvin Demke and Seraphin Bücklers for their support and helpful discussions.
The HK Project v1995 NSO data and their v2001 extension derive from the Mount Wilson Observatory HK Project, which was supported by both public and private funds through the Carnegie Observatories, the Mount Wilson Institute, and the Harvard-Smithsonian Center for Astrophysics starting in 1966 and continuing for over 36 years. These data are the result of the dedicated work of O. Wilson, A. Vaughan, G. Preston, D. Duncan, S. Baliunas, and many others.

We would like to thank the staff at Lick Observatory for their support and thank the CAT observers who assisted with this project, including David Mitchell, Saskia Hekker, Christian Schwab, Christoph Bergmann, Kirsten Vincke, Julian Stürmer, Kelsey Clubb, David Bauer, Dominika Wylezalek, Dennis Kügler, and Debra Fischer.

This research has made use of the VizieR catalogue access tool and the SIMBAD database, operated at CDS, Strasbourg, France. 

This research made use of Astropy,\footnote{\url{http://www.astropy.org}} a community-developed core Python package for Astronomy \citep{astropy2013, astropy2018, astropy2022}. We further acknowledge use of the Python packages NumPy \citep{harris2020}, SciPy \citep{virtanen2020}, Matplotlib \citep{Hunter2007}, pandas \citep{thepandasdevelopmentteam2026, McKinney2010}, PyAstronomy\footnote{\url{https://github.com/sczesla/PyAstronomy}} \citep{Czesla2019}, \texttt{celerite} \citep{foreman-mackey2017}, SOAP \citep{boisse2012,cristo2025}, and \texttt{emcee} \citep{foreman-mackey2013}.

This work has made use of data from the European Space Agency (ESA) mission \textsc{Hipparcos} \citep{perryman1997}. We made use of photometry from the MASCARA survey \citep{talens2017, burggraaff2018} and of the HIDES radial velocities of $\varepsilon$ Psc published by \citet{teng2022}.

\end{acknowledgements}

\bibliography{Rotationpaper}

\begin{thebibliography}{98}
\expandafter\ifx\csname natexlab\endcsname\relax\def\natexlab#1{#1}\fi

\bibitem[{Auri{\`e}re {et~al.}(2015)Auri{\`e}re, {Konstantinova-Antova},
  Charbonnel, Wade, Tsvetkova, Petit, Dintrans, Drake, Decressin, Lagarde,
  Donati, Roudier, Ligni{\`e}res, Schr{\"o}der, Landstreet, L{\`e}bre, Weiss,
  \& Zahn}]{auriere2015}
Auri{\`e}re, M., {Konstantinova-Antova}, R., Charbonnel, C., {et~al.} 2015,
  \aap, 574, A90

\bibitem[{Auri{\`e}re {et~al.}(2014)Auri{\`e}re, {Konstantinova-Antova},
  Espagnet, Petit, Roudier, Charbonnel, Donati, \& Wade}]{auriere2013}
Auri{\`e}re, M., {Konstantinova-Antova}, R., Espagnet, O., {et~al.} 2014, in
  IAU Symposium, Vol. 302, Magnetic Fields throughout Stellar Evolution, ed.
  P.~Petit, M.~Jardine, \& H.~C. Spruit, 359

\bibitem[{Baliunas {et~al.}(1998)Baliunas, Donahue, Soon, \&
  Henry}]{baliunas1998}
Baliunas, S.~L., Donahue, R.~A., Soon, W., \& Henry, G.~W. 1998, in ASP Conf.
  Ser., Vol. 154, Cool {{Stars}}, {{Stellar Systems}}, and the {{Sun}}, ed.
  R.~A. Donahue \& J.~A. Bookbinder, 153

\bibitem[{Baliunas {et~al.}(1995)Baliunas, Donahue, Soon, Horne, Frazer,
  {Woodard-Eklund}, Bradford, Rao, Wilson, Zhang, Bennett, Briggs, Carroll,
  Duncan, Figueroa, Lanning, Misch, Mueller, Noyes, Poppe, Porter, Robinson,
  Russell, Shelton, Soyumer, Vaughan, \& Whitney}]{baliunas1995a}
Baliunas, S.~L., Donahue, R.~A., Soon, W.~H., {et~al.} 1995, \apj, 438, 269

\bibitem[{Baliunas {et~al.}(1983)Baliunas, Hartmann, Noyes, Vaughan, Preston,
  Frazer, Lanning, Middelkoop, \& Mihalas}]{baliunas1983}
Baliunas, S.~L., Hartmann, L., Noyes, R.~W., {et~al.} 1983, \apj, 275, 752

\bibitem[{Baliunas {et~al.}(1985)Baliunas, Horne, Porter, Duncan, Frazer,
  Lanning, Misch, Mueller, Noyes, Soyumer, Vaughan, \& Woodard}]{baliunas1985}
Baliunas, S.~L., Horne, J.~H., Porter, A., {et~al.} 1985, \apj, 294, 310

\bibitem[{Barnes(2003)}]{barnes2003}
Barnes, S.~A. 2003, \apj, 586, 464

\bibitem[{Berdyugina(2005)}]{berdyugina2005}
Berdyugina, S.~V. 2005, Living Rev. Sol. Phys., 2, 8

\bibitem[{Boisse {et~al.}(2012)Boisse, Bonfils, \& Santos}]{boisse2012}
Boisse, I., Bonfils, X., \& Santos, N.~C. 2012, \aap, 545, A109

\bibitem[{Boisse {et~al.}(2011)Boisse, Bouchy, H{\'e}brard, Bonfils, Santos, \&
  Vauclair}]{boisse2011}
Boisse, I., Bouchy, F., H{\'e}brard, G., {et~al.} 2011, \aap, 528, A4

\bibitem[{Burggraaff {et~al.}(2018)Burggraaff, Talens, Spronck, Lesage, Stuik,
  Otten, Van~Eylen, Pollacco, \& Snellen}]{burggraaff2018}
Burggraaff, O., Talens, G. J.~J., Spronck, J., {et~al.} 2018, \aap, 617, A32

\bibitem[{Butler {et~al.}(1996)Butler, Marcy, Williams, McCarthy, Dosanjh, \&
  Vogt}]{butler1996}
Butler, R.~P., Marcy, G.~W., Williams, E., {et~al.} 1996, \pasp, 108, 500

\bibitem[{Byrd {et~al.}(1995)Byrd, Lu, Nocedal, \& Zhu}]{byrd1995}
Byrd, R.~H., Lu, P., Nocedal, J., \& Zhu, C. 1995, SIAM J. Sci. Comput., 16,
  1190

\bibitem[{Ceillier {et~al.}(2017)Ceillier, Tayar, Mathur, Salabert, Garc{\'i}a,
  Stello, Pinsonneault, Van~Saders, Beck, \& Bloemen}]{ceillier2017}
Ceillier, T., Tayar, J., Mathur, S., {et~al.} 2017, \aap, 605, A111

\bibitem[{Choi {et~al.}(1995)Choi, Soon, Donahue, Baliunas, \&
  Henry}]{choi1995}
Choi, H.-J., Soon, W., Donahue, R.~A., Baliunas, S.~L., \& Henry, G.~W. 1995,
  \pasp, 107, 744

\bibitem[{Cretignier {et~al.}(2023)Cretignier, Pietrow, \&
  Aigrain}]{cretignier2023a}
Cretignier, M., Pietrow, A. G.~M., \& Aigrain, S. 2023, \mnras, 527, 2940

\bibitem[{Cristo {et~al.}(2025)Cristo, Faria, Santos, Dethier, Akinsanmi,
  Barka, Demangeon, Lucero, \& Silva}]{cristo2025}
Cristo, E., Faria, J.~P., Santos, N.~C., {et~al.} 2025, \aap, 702, A84

\bibitem[{Czesla {et~al.}(2019)Czesla, Schr{\"o}ter, Schneider, Huber, Pfeifer,
  Andreasen, \& Zechmeister}]{Czesla2019}
Czesla, S., Schr{\"o}ter, S., Schneider, C.~P., {et~al.} 2019, Astrophysics
  Source Code Library [\eprint{ascl:1906.010}]

\bibitem[{De~Medeiros {et~al.}(2014)De~Medeiros, Alves, Udry, Andersen,
  Nordstr{\"o}m, \& Mayor}]{demedeiros2014}
De~Medeiros, J.~R., Alves, S., Udry, S., {et~al.} 2014, \aap, 561, A126

\bibitem[{De~Medeiros \& Mayor(1999)}]{demedeiros1999}
De~Medeiros, J.~R. \& Mayor, M. 1999, \aaps, 139, 433

\bibitem[{Dixon {et~al.}(2020)Dixon, Tayar, \& Stassun}]{dixon2020}
Dixon, D., Tayar, J., \& Stassun, K.~G. 2020, \aj, 160, 12

\bibitem[{Donahue {et~al.}(1996)Donahue, Saar, \& Baliunas}]{donahue1996}
Donahue, R.~A., Saar, S.~H., \& Baliunas, S.~L. 1996, \apj, 466, 384

\bibitem[{Dumusque {et~al.}(2014)Dumusque, Boisse, \& Santos}]{dumusque2014}
Dumusque, X., Boisse, I., \& Santos, N.~C. 2014, \apj, 796, 132

\bibitem[{Duncan {et~al.}(1991)Duncan, Vaughan, Wilson, Preston, Frazer,
  Lanning, Misch, Mueller, Soyumer, Woodard, Baliunas, Noyes, Hartmann, Porter,
  Zwaan, Middelkoop, Rutten, \& Mihalas}]{duncan1991}
Duncan, D.~K., Vaughan, A.~H., Wilson, O.~C., {et~al.} 1991, \apjs, 76, 383

\bibitem[{Fekel {et~al.}(1986)Fekel, Moffett, \& Henry}]{fekel1986}
Fekel, F.~C., Moffett, T.~J., \& Henry, G.~W. 1986, \apjs, 60, 551

\bibitem[{{Foreman-Mackey} {et~al.}(2017){Foreman-Mackey}, Agol, Ambikasaran,
  \& Angus}]{foreman-mackey2017}
{Foreman-Mackey}, D., Agol, E., Ambikasaran, S., \& Angus, R. 2017, \aj, 154,
  220

\bibitem[{{Foreman-Mackey} {et~al.}(2013){Foreman-Mackey}, Hogg, Lang, \&
  Goodman}]{foreman-mackey2013}
{Foreman-Mackey}, D., Hogg, D.~W., Lang, D., \& Goodman, J. 2013, \pasp, 125,
  306

\bibitem[{Frandsen {et~al.}(2002)Frandsen, Carrier, Aerts, Stello, Maas,
  Burnet, Bruntt, Teixeira, De~Medeiros, Bouchy, Kjeldsen, Pijpers, \&
  {Christensen-Dalsgaard}}]{frandsen2002}
Frandsen, S., Carrier, F., Aerts, C., {et~al.} 2002, \aap, 394, L5

\bibitem[{Frink {et~al.}(2002)Frink, Mitchell, Quirrenbach, Fischer, Marcy, \&
  Butler}]{frink2002}
Frink, S., Mitchell, D.~S., Quirrenbach, A., {et~al.} 2002, \apj, 576, 478

\bibitem[{Frink {et~al.}(2001)Frink, Quirrenbach, Fischer, R{\"o}ser, \&
  Schilbach}]{frink2001}
Frink, S., Quirrenbach, A., Fischer, D., R{\"o}ser, S., \& Schilbach, E. 2001,
  \pasp, 113, 173

\bibitem[{Fuhrmeister {et~al.}(2019)Fuhrmeister, Czesla, Schmitt, Johnson,
  Sch{\"o}fer, Jeffers, Caballero, Zechmeister, Reiners, Ribas, Amado,
  Quirrenbach, Bauer, B{\'e}jar, {Cort{\'e}s-Contreras}, D{\'i}ez~Alonso,
  Dreizler, {Galad{\'i}-Enr{\'i}quez}, Guenther, Kaminski, K{\"u}rster,
  Lafarga, \& Montes}]{fuhrmeister2019}
Fuhrmeister, B., Czesla, S., Schmitt, J. H. M.~M., {et~al.} 2019, \aap, 623,
  A24

\bibitem[{Giles {et~al.}(2017)Giles, Collier~Cameron, \& Haywood}]{giles2017}
Giles, H. A.~C., Collier~Cameron, A., \& Haywood, R.~D. 2017, \mnras, 472, 1618

\bibitem[{Gilliland \& Fisher(1985)}]{gilliland1985}
Gilliland, R.~L. \& Fisher, R. 1985, \pasp, 97, 285

\bibitem[{Golovin {et~al.}(2023)Golovin, Reffert, Just, Jordan, Vani, \&
  Jahrei{\ss}}]{golovin2023}
Golovin, A., Reffert, S., Just, A., {et~al.} 2023, \aap, 670, A19

\bibitem[{Gray(1989)}]{gray1989}
Gray, D.~F. 1989, \apj, 347, 1021

\bibitem[{Gray(2013)}]{gray2013}
Gray, D.~F. 2013, \aj, 146, 29

\bibitem[{Harris {et~al.}(2020)Harris, Millman, Van Der~Walt, Gommers,
  Virtanen, Cournapeau, Wieser, Taylor, Berg, Smith, Kern, Picus, Hoyer,
  Van~Kerkwijk, Brett, Haldane, Del~R{\'i}o, Wiebe, Peterson,
  {G{\'e}rard-Marchant}, Sheppard, Reddy, Weckesser, Abbasi, Gohlke, \&
  Oliphant}]{harris2020}
Harris, C.~R., Millman, K.~J., Van Der~Walt, S.~J., {et~al.} 2020, \nat, 585,
  357

\bibitem[{Hatzes \& Cochran(1996)}]{hatzes1996}
Hatzes, A.~P. \& Cochran, W.~D. 1996, \apj, 468, 391

\bibitem[{Hatzes \& Cochran(1999)}]{hatzes1999}
Hatzes, A.~P. \& Cochran, W.~D. 1999, \mnras, 304, 109

\bibitem[{Hatzes {et~al.}(2015)Hatzes, Cochran, Endl, Guenther, MacQueen,
  Hartmann, Zechmeister, Han, Lee, Walker, Yang, Larson, Kim, Mkrtichian,
  D{\"o}llinger, Simon, \& Girardi}]{hatzes2015}
Hatzes, A.~P., Cochran, W.~D., Endl, M., {et~al.} 2015, \aap, 580, A31

\bibitem[{Heeren {et~al.}(2021)Heeren, Reffert, Trifonov, Wong, Lee,
  {Lillo-Box}, Quirrenbach, Arentoft, Albrecht, Grundahl, Andersen, Antoci, \&
  Pall{\'e}}]{heeren2021}
Heeren, P., Reffert, S., Trifonov, T., {et~al.} 2021, \aap, 647, A160

\bibitem[{Hekker \& Mel{\'e}ndez(2007)}]{hekker2007}
Hekker, S. \& Mel{\'e}ndez, J. 2007, \aap, 475, 1003

\bibitem[{Hekker {et~al.}(2008)Hekker, Snellen, Aerts, Quirrenbach, Reffert, \&
  Mitchell}]{hekker2008}
Hekker, S., Snellen, I. A.~G., Aerts, C., {et~al.} 2008, \aap, 480, 215

\bibitem[{Henry {et~al.}(2000)Henry, Fekel, Henry, \& Hall}]{henry2000}
Henry, G.~W., Fekel, F.~C., Henry, S.~M., \& Hall, D.~S. 2000, \apjs, 130, 201

\bibitem[{Hojjatpanah {et~al.}(2019)Hojjatpanah, Figueira, Santos, Adibekyan,
  Sousa, {Delgado-Mena}, Alibert, Cristiani, Gonz{\'a}lez~Hern{\'a}ndez, Lanza,
  Di~Marcantonio, Martins, Micela, Molaro, Neves, Oshagh, Pepe, Poretti,
  {Rojas-Ayala}, Rebolo, Su{\'a}rez~Mascare{\~n}o, \&
  Zapatero~Osorio}]{hojjatpanah2019}
Hojjatpanah, S., Figueira, P., Santos, N.~C., {et~al.} 2019, \aap, 629, A80

\bibitem[{Hopkins {et~al.}(1985)Hopkins, Boyd, Genet, \& Hall}]{hopkins1985}
Hopkins, J.~L., Boyd, L.~J., Genet, R.~M., \& Hall, D.~S. 1985, IBVS, 2684, 1

\bibitem[{Hunter(2007)}]{Hunter2007}
Hunter, J.~D. 2007, Comput. Sci. Eng., 9, 90

\bibitem[{Hussain(2002)}]{hussain2002}
Hussain, G. 2002, Astron. Nachr., 323, 349

\bibitem[{Jofr{\'e} {et~al.}(2015)Jofr{\'e}, Petrucci, Saffe, Saker, Artur De
  La~Villarmois, Chavero, G{\'o}mez, \& Mauas}]{jofre2015}
Jofr{\'e}, E., Petrucci, R., Saffe, C., {et~al.} 2015, \aap, 574, A50

\bibitem[{J{\"o}nsson {et~al.}(2020)J{\"o}nsson, Holtzman, Prieto, Cunha,
  {Garc{\'i}a-Hern{\'a}ndez}, Hasselquist, Masseron, Osorio, Shetrone, Smith,
  Stringfellow, Bizyaev, Edvardsson, Majewski, M{\'e}sz{\'a}ros, Souto, Zamora,
  Beaton, Bovy, Donor, Pinsonneault, Poovelil, \& Sobeck}]{jonsson2020}
J{\"o}nsson, H., Holtzman, J.~A., Prieto, C.~A., {et~al.} 2020, \aj, 160, 120

\bibitem[{Kjeldsen \& Bedding(1995)}]{kjeldsen1995}
Kjeldsen, H. \& Bedding, T.~R. 1995, \aap, 293, 87

\bibitem[{Koen \& Eyer(2002)}]{koen2002}
Koen, C. \& Eyer, L. 2002, \mnras, 331, 45

\bibitem[{Lafarga {et~al.}(2021)Lafarga, Ribas, Reiners, Quirrenbach, Amado,
  Caballero, Azzaro, B{\'e}jar, {Cort{\'e}s-Contreras}, Dreizler, Hatzes,
  Henning, Jeffers, Kaminski, K{\"u}rster, Montes, Morales, Oshagh,
  {Rodr{\'i}guez-L{\'o}pez}, Sch{\"o}fer, Schweitzer, \&
  Zechmeister}]{lafarga2021}
Lafarga, M., Ribas, I., Reiners, A., {et~al.} 2021, \aap, 652, A28

\bibitem[{Lee {et~al.}(2014)Lee, Han, Park, Mkrtichian, Jeong, Kim, \&
  Valyavin}]{lee2014}
Lee, B.-C., Han, I., Park, M.-G., {et~al.} 2014, J. Korean Astron. Soc., 47, 69

\bibitem[{Lehtinen {et~al.}(2020)Lehtinen, Spada, K{\"a}pyl{\"a}, Olspert, \&
  K{\"a}pyl{\"a}}]{lehtinen2020}
Lehtinen, J.~J., Spada, F., K{\"a}pyl{\"a}, M.~J., Olspert, N., \&
  K{\"a}pyl{\"a}, P.~J. 2020, Nat. Astron., 4, 658

\bibitem[{Liebing {et~al.}(2023)Liebing, Jeffers, Zechmeister, \&
  Reiners}]{liebing2023}
Liebing, F., Jeffers, S.~V., Zechmeister, M., \& Reiners, A. 2023, \aap, 673,
  A43

\bibitem[{Massarotti {et~al.}(2008)Massarotti, Latham, Stefanik, \&
  Fogel}]{massarotti2008}
Massarotti, A., Latham, D.~W., Stefanik, R.~P., \& Fogel, J. 2008, \aj, 135,
  209

\bibitem[{Mayor \& Queloz(1995)}]{mayor1995}
Mayor, M. \& Queloz, D. 1995, \nat, 378, 355

\bibitem[{McKinney(2010)}]{McKinney2010}
McKinney, W. 2010, in Proceedings of the 9th {{Python}} in {{Science
  Conference}}, ed. S.~{van der Walt} \& J.~Millman, 56--61

\bibitem[{Mortier \& Collier~Cameron(2017)}]{mortier2017}
Mortier, A. \& Collier~Cameron, A. 2017, \aap, 601, A110

\bibitem[{Noyes {et~al.}(1984)Noyes, Hartmann, Baliunas, Duncan, \&
  Vaughan}]{noyes1984}
Noyes, R.~W., Hartmann, L.~W., Baliunas, S.~L., Duncan, D.~K., \& Vaughan,
  A.~H. 1984, \apj, 279, 763

\bibitem[{Pasquini {et~al.}(2000)Pasquini, {de Medeiros}, \&
  Girardi}]{pasquini2000}
Pasquini, L., {de Medeiros}, J.~R., \& Girardi, L. 2000, \aap, 361, 1011

\bibitem[{Perryman {et~al.}(1997)Perryman, Lindegren, Kovalevsky, Hoeg,
  Bastian, Bernacca, Cr{\'e}z{\'e}, Donati, Grenon, Grewing, {van Leeuwen},
  {van der Marel}, Mignard, Murray, Le~Poole, Schrijver, Turon, Arenou,
  Froeschl{\'e}, \& Petersen}]{perryman1997}
Perryman, M. A.~C., Lindegren, L., Kovalevsky, J., {et~al.} 1997, \aap, 323,
  L49

\bibitem[{Quirrenbach {et~al.}(2019)Quirrenbach, Trifonov, Lee, \&
  Reffert}]{quirrenbach2019}
Quirrenbach, A., Trifonov, T., Lee, M.~H., \& Reffert, S. 2019, \aap, 624, A18

\bibitem[{Radick \& Pevtsov(2018{\natexlab{a}})}]{radick2018}
Radick, R. \& Pevtsov, A. 2018{\natexlab{a}},
  {{HK}}\_{{Project}}\_v1995\_{{NSO}}, Harvard Dataverse,
  https://doi.org/10.7910/DVN/ZRJ6NT

\bibitem[{Radick \& Pevtsov(2018{\natexlab{b}})}]{radick2018a}
Radick, R. \& Pevtsov, A. 2018{\natexlab{b}},
  {{HK}}\_{{Project}}\_v2001\_{{NSO}}, Harvard Dataverse,
  https://doi.org/10.7910/DVN/H85QPF

\bibitem[{Reffert {et~al.}(2015)Reffert, Bergmann, Quirrenbach, Trifonov, \&
  K{\"u}nstler}]{reffert2015}
Reffert, S., Bergmann, C., Quirrenbach, A., Trifonov, T., \& K{\"u}nstler, A.
  2015, \aap, 574, A116

\bibitem[{Reichert {et~al.}(2019)Reichert, Reffert, Stock, Trifonov, \&
  Quirrenbach}]{reichert2019}
Reichert, K., Reffert, S., Stock, S., Trifonov, T., \& Quirrenbach, A. 2019,
  \aap, 625, A22

\bibitem[{Rutten(1984)}]{rutten1984}
Rutten, R. G.~M. 1984, \aap, 130, 353

\bibitem[{Saar \& Donahue(1997)}]{saar1997}
Saar, S.~H. \& Donahue, R.~A. 1997, \apj, 485, 319

\bibitem[{Saio {et~al.}(2015)Saio, Wood, Takayama, \& Ita}]{saio2015}
Saio, H., Wood, P.~R., Takayama, M., \& Ita, Y. 2015, \mnras, 452, 3863

\bibitem[{Schr{\"o}der {et~al.}(2020)Schr{\"o}der, Mittag, Jack,
  Rodr{\'i}guez~Jim{\'e}nez, \& Schmitt}]{schroder2020}
Schr{\"o}der, K.-P., Mittag, M., Jack, D., Rodr{\'i}guez~Jim{\'e}nez, A., \&
  Schmitt, J. H. M.~M. 2020, \mnras, 492, 1110

\bibitem[{Skumanich(1972)}]{skumanich1972}
Skumanich, A. 1972, \apj, 171, 565

\bibitem[{Sowmya {et~al.}(2023)Sowmya, Shapiro, Rouppe Van Der~Voort, Krivova,
  \& Solanki}]{sowmya2023}
Sowmya, K., Shapiro, A.~I., Rouppe Van Der~Voort, L. H.~M., Krivova, N.~A., \&
  Solanki, S.~K. 2023, \apjl, 956, L10

\bibitem[{Spaeth {et~al.}(2024)Spaeth, Reffert, Hunt, Kaminski, \&
  Quirrenbach}]{spaeth2024}
Spaeth, D., Reffert, S., Hunt, E.~L., Kaminski, A., \& Quirrenbach, A. 2024,
  \aap, 689, A91

\bibitem[{Spaeth {et~al.}(2025)Spaeth, Reffert, Trifonov, Kaminski, Albrecht,
  Grundahl, Fredslund~Andersen, Quirrenbach, \& Pall{\'e}}]{spaeth2025}
Spaeth, D., Reffert, S., Trifonov, T., {et~al.} 2025, \aap, 697, A32

\bibitem[{Stock {et~al.}(2018)Stock, Reffert, \& Quirrenbach}]{stock2018}
Stock, S., Reffert, S., \& Quirrenbach, A. 2018, \aap, 616, A33

\bibitem[{Strassmeier \& Hall(1988)}]{strassmeier1988}
Strassmeier, K. \& Hall, D.~S. 1988, \apjs, 67, 453

\bibitem[{Strassmeier {et~al.}(1999)Strassmeier, Serkowitsch, \&
  Granzer}]{strassmeier1999}
Strassmeier, K.~G., Serkowitsch, E., \& Granzer, {\relax Th}. 1999, \aaps, 140,
  29

\bibitem[{Su{\'a}rez~Mascare{\~n}o {et~al.}(2015)Su{\'a}rez~Mascare{\~n}o,
  Rebolo, Gonz{\'a}lez~Hern{\'a}ndez, \& Esposito}]{suarezmascareno2015}
Su{\'a}rez~Mascare{\~n}o, A., Rebolo, R., Gonz{\'a}lez~Hern{\'a}ndez, J.~I., \&
  Esposito, M. 2015, \mnras, 452, 2745

\bibitem[{Takeda {et~al.}(2008)Takeda, Sato, \& Murata}]{takeda2008}
Takeda, Y., Sato, B., \& Murata, D. 2008, \pasj, 60, 781

\bibitem[{Talens {et~al.}(2017)Talens, Spronck, Lesage, Otten, Stuik, Pollacco,
  \& Snellen}]{talens2017}
Talens, G. J.~J., Spronck, J. F.~P., Lesage, A.-L., {et~al.} 2017, \aap, 601,
  A11

\bibitem[{Tayar {et~al.}(2019)Tayar, Stassun, \& Corsaro}]{tayar2019}
Tayar, J., Stassun, K.~G., \& Corsaro, E. 2019, \apj, 883, 195

\bibitem[{Teng {et~al.}(2022)Teng, Sato, Takarada, Omiya, Harakawa, Izumiura,
  Kambe, Takeda, Yoshida, Itoh, Ando, \& Kokubo}]{teng2022}
Teng, H.-Y., Sato, B., Takarada, T., {et~al.} 2022, \pasj, 74, 92

\bibitem[{{The Astropy Collaboration} {et~al.}(2022){The Astropy
  Collaboration}, {Price-Whelan}, Lim, Earl, Starkman, Bradley, Shupe, Patil,
  Corrales, Brasseur, N{\"o}the, Donath, Tollerud, Morris, Ginsburg, Vaher,
  Weaver, Tocknell, Jamieson, Van~Kerkwijk, Robitaille, Merry, Bachetti,
  G{\"u}nther, {Paper Authors}, Aldcroft, {Alvarado-Montes}, Archibald,
  B{\'o}di, Bapat, Barentsen, Baz{\'a}n, Biswas, Boquien, Burke, Cara, Cara,
  Conroy, Conseil, Craig, Cross, Cruz, D'Eugenio, Dencheva, Devillepoix,
  Dietrich, Eigenbrot, Erben, Ferreira, {Foreman-Mackey}, Fox, Freij, Garg,
  Geda, Glattly, Gondhalekar, Gordon, Grant, Greenfield, Groener, Guest,
  Gurovich, Handberg, Hart, {Hatfield-Dodds}, Homeier, Hosseinzadeh, Jenness,
  Jones, Joseph, Kalmbach, Karamehmetoglu, Ka{\l}uszy{\'n}ski, Kelley, Kern,
  Kerzendorf, Koch, Kulumani, Lee, Ly, Ma, MacBride, Maljaars, Muna, Murphy,
  Norman, O'Steen, Oman, Pacifici, Pascual, {Pascual-Granado}, Patil, Perren,
  Pickering, Rastogi, Roulston, Ryan, Rykoff, Sabater, Sakurikar, Salgado,
  Sanghi, Saunders, Savchenko, Schwardt, {Seifert-Eckert}, Shih, Jain, Shukla,
  Sick, Simpson, Singanamalla, Singer, Singhal, Sinha, Sip{\H o}cz, Spitler,
  Stansby, Streicher, {\v S}umak, Swinbank, Taranu, Tewary, Tremblay,
  {Val-Borro}, Van~Kooten, Vasovi{\'c}, Verma, De~Miranda~Cardoso, Williams,
  Wilson, Winkel, {Wood-Vasey}, Xue, Yoachim, Zhang, Zonca, \& {Astropy Project
  Contributors}}]{astropy2022}
{The Astropy Collaboration}, {Price-Whelan}, A.~M., Lim, P.~L., {et~al.} 2022,
  \apj, 935, 167

\bibitem[{{The Astropy Collaboration} {et~al.}(2018){The Astropy
  Collaboration}, {Price-Whelan}, Sip{\H o}cz, G{\"u}nther, Lim, Crawford,
  Conseil, Shupe, Craig, Dencheva, Ginsburg, VanderPlas, Bradley,
  {P{\'e}rez-Su{\'a}rez}, {De Val-Borro}, {(Primary Paper Contributors)},
  Aldcroft, Cruz, Robitaille, Tollerud, {(Astropy Coordination Committee)},
  Ardelean, Babej, Bach, Bachetti, Bakanov, Bamford, Barentsen, Barmby,
  Baumbach, Berry, Biscani, Boquien, Bostroem, Bouma, Brammer, Bray,
  Breytenbach, Buddelmeijer, Burke, Calderone, Rodr{\'i}guez, Cara, Cardoso,
  Cheedella, Copin, Corrales, Crichton, D'Avella, Deil, Depagne, Dietrich,
  Donath, Droettboom, Earl, Erben, Fabbro, Ferreira, Finethy, Fox, Garrison,
  Gibbons, Goldstein, Gommers, Greco, Greenfield, Groener, Grollier, Hagen,
  Hirst, Homeier, Horton, Hosseinzadeh, Hu, Hunkeler, Ivezi{\'c}, Jain,
  Jenness, Kanarek, Kendrew, Kern, Kerzendorf, Khvalko, King, Kirkby, Kulkarni,
  Kumar, Lee, Lenz, Littlefair, Ma, Macleod, Mastropietro, McCully, Montagnac,
  Morris, Mueller, Mumford, Muna, Murphy, Nelson, Nguyen, Ninan, N{\"o}the,
  Ogaz, Oh, Parejko, Parley, Pascual, Patil, Patil, Plunkett, Prochaska,
  Rastogi, Janga, Sabater, Sakurikar, Seifert, Sherbert, {Sherwood-Taylor},
  Shih, Sick, Silbiger, Singanamalla, Singer, Sladen, Sooley, Sornarajah,
  Streicher, Teuben, Thomas, Tremblay, Turner, Terr{\'o}n, Kerkwijk,
  De~La~Vega, Watkins, Weaver, Whitmore, Woillez, Zabalza, \& {(Astropy
  Contributors)}}]{astropy2018}
{The Astropy Collaboration}, {Price-Whelan}, A.~M., Sip{\H o}cz, B.~M.,
  {et~al.} 2018, \aj, 156, 123

\bibitem[{{The Astropy Collaboration} {et~al.}(2013){The Astropy
  Collaboration}, Robitaille, Tollerud, Greenfield, Droettboom, Bray, Aldcroft,
  Davis, Ginsburg, {Price-Whelan}, Kerzendorf, Conley, Crighton, Barbary, Muna,
  Ferguson, Grollier, Parikh, Nair, G{\"u}nther, Deil, Woillez, Conseil,
  Kramer, Turner, Singer, Fox, Weaver, Zabalza, Edwards, Azalee~Bostroem,
  Burke, Casey, Crawford, Dencheva, Ely, Jenness, Labrie, Lim, Pierfederici,
  Pontzen, Ptak, Refsdal, Servillat, \& Streicher}]{astropy2013}
{The Astropy Collaboration}, Robitaille, T.~P., Tollerud, E.~J., {et~al.} 2013,
  \aap, 558, A33

\bibitem[{{The pandas development team}(2026)}]{thepandasdevelopmentteam2026}
{The pandas development team}. 2026, {pandas-dev/pandas: Pandas}, Zenodo,
  https://doi.org/10.5281/zenodo.3509134

\bibitem[{Trifonov {et~al.}(2014)Trifonov, Reffert, Tan, Lee, \&
  Quirrenbach}]{trifonov2014}
Trifonov, T., Reffert, S., Tan, X., Lee, M.~H., \& Quirrenbach, A. 2014, \aap,
  568, A64

\bibitem[{Tsvetkova {et~al.}(2013)Tsvetkova, Petit, Auri{\`e}re,
  {Konstantinova-Antova}, Wade, Charbonnel, Decressin, \&
  Bogdanovski}]{tsvetkova2013}
Tsvetkova, S., Petit, P., Auri{\`e}re, M., {et~al.} 2013, \aap, 556, A43

\bibitem[{{van Leeuwen} {et~al.}(1997){van Leeuwen}, Evans, Grenon, Grossmann,
  Mignard, \& Perryman}]{vanleeuwen1997}
{van Leeuwen}, F., Evans, D.~W., Grenon, M., {et~al.} 1997, \aap, 323, L61

\bibitem[{Vanderburg {et~al.}(2016)Vanderburg, Plavchan, Johnson, Ciardi,
  Swift, \& Kane}]{vanderburg2016}
Vanderburg, A., Plavchan, P., Johnson, J.~A., {et~al.} 2016, \mnras, 459, 3565

\bibitem[{Virtanen {et~al.}(2020)Virtanen, Gommers, Oliphant, Haberland, Reddy,
  Cournapeau, Burovski, Peterson, Weckesser, Bright, Van Der~Walt, Brett,
  Wilson, Millman, Mayorov, Nelson, Jones, Kern, Larson, Carey, Polat, Feng,
  Moore, VanderPlas, Laxalde, Perktold, Cimrman, Henriksen, Quintero, Harris,
  Archibald, Ribeiro, Pedregosa, Van~Mulbregt, {SciPy 1.0 Contributors},
  Vijaykumar, Bardelli, Rothberg, Hilboll, Kloeckner, Scopatz, Lee, Rokem,
  Woods, Fulton, Masson, H{\"a}ggstr{\"o}m, Fitzgerald, Nicholson, Hagen,
  Pasechnik, Olivetti, Martin, Wieser, Silva, Lenders, Wilhelm, Young, Price,
  Ingold, Allen, Lee, Audren, Probst, Dietrich, Silterra, Webber, Slavi{\v c},
  Nothman, Buchner, Kulick, Sch{\"o}nberger, De~Miranda~Cardoso, Reimer,
  Harrington, Rodr{\'i}guez, {Nunez-Iglesias}, Kuczynski, Tritz, Thoma,
  Newville, K{\"u}mmerer, Bolingbroke, Tartre, Pak, Smith, Nowaczyk, Shebanov,
  Pavlyk, Brodtkorb, Lee, McGibbon, Feldbauer, Lewis, Tygier, Sievert, Vigna,
  Peterson, More, Pudlik, Oshima, Pingel, Robitaille, Spura, Jones, Cera,
  Leslie, Zito, Krauss, Upadhyay, Halchenko, \&
  {V{\'a}zquez-Baeza}}]{virtanen2020}
Virtanen, P., Gommers, R., Oliphant, T.~E., {et~al.} 2020, Nat. Methods, 17,
  261

\bibitem[{Vogt(1987)}]{vogt1987}
Vogt, S.~S. 1987, \pasp, 99, 1214

\bibitem[{Wilson(1968)}]{wilson1968}
Wilson, O.~C. 1968, \apj, 153, 221

\bibitem[{Zechmeister \& K{\"u}rster(2009)}]{zechmeister2009}
Zechmeister, M. \& K{\"u}rster, M. 2009, \aap, 496, 577

\bibitem[{Zechmeister {et~al.}(2018)Zechmeister, Reiners, Amado, Azzaro, Bauer,
  B{\'e}jar, Caballero, Guenther, Hagen, Jeffers, Kaminski, K{\"u}rster,
  Launhardt, Montes, Morales, Quirrenbach, Reffert, Ribas, Seifert, {Tal-Or},
  \& Wolthoff}]{zechmeister2018}
Zechmeister, M., Reiners, A., Amado, P.~J., {et~al.} 2018, \aap, 609, A12

\bibitem[{Zhao {et~al.}(2022)Zhao, Fischer, Ford, Wise, Cretignier, Aigrain,
  Barragan, Bedell, Buchhave, Camacho, Cegla, {Cisewski-Kehe}, Collier~Cameron,
  De~Beurs, {Dodson-Robinson}, Dumusque, Faria, Gilbertson, Haley, Harrell,
  Hogg, Holzer, John, Klein, Lafarga, Lienhard, {Maguire-Rajpaul}, Mortier,
  Nicholson, Palumbo, Ramirez~Delgado, Shallue, Vanderburg, Viana, Zhao,
  Zicher, Cabot, Henry, Roettenbacher, Brewer, Llama, Petersburg, \&
  Szymkowiak}]{zhao2022}
Zhao, L.~L., Fischer, D.~A., Ford, E.~B., {et~al.} 2022, \aj, 163, 171

\end{thebibliography}
\newpage
\begin{appendix}
\nolinenumbers

\onecolumn

\section{Significant periods and stellar parameters}
\label{app:results_table}
\begin{longtable}{lllrrrrrrrrrrl}
\caption{Significant periods and stellar parameters.}\\\\
\label{tab:results}\\
\hline\hline
HIP & HD & Name & $P_\mathrm{cyc}$ & $P_\mathrm{rot}$ & $P_\mathrm{S}$ & $P_\mathrm{phot}$ & $P_\mathrm{rv}$ & $P_\mathrm{lit}$ & $P_\mathrm{max}$ & $v\sin i$ & $R$ & Stage & Notes \\
 & & & (d) & (d) & (d) & (d) & (d) & (d) & (d) & (km\,s$^{-1}$) & ($R_\odot$) & & \\
\hline
\endfirsthead
\caption{continued.}\\
\hline\hline
HIP & HD & Name & $P_\mathrm{cyc}$ & $P_\mathrm{rot}$ & $P_\mathrm{S}$ & $P_\mathrm{phot}$ & $P_\mathrm{rv}$ & $P_\mathrm{lit}$ & $P_\mathrm{max}$ & $v\sin i$ & $R$ & Stage & Notes \\
 & & & (d) & (d) & (d) & (d) & (d) & (d) & (d) & (km\,s$^{-1}$) & ($R_\odot$) & & \\
\hline
\endhead
\hline
\endfoot
1562 & 1522 & $\iota$ Cet & \ldots & 781 & 809 & 781 & \ldots & \ldots & 601 & 2.6\rlap{\,$^{(4)}$}\hphantom{\,$^{(11)}$} & 31.0 & HB &  \\
 &  &  &  & $62$ & $37$ & $62$ &  &  & $^{+417}_{-177}$ & $1.1$\hphantom{\,$^{(11)}$} & $^{+0.5}_{-0.5}$ &  &  \\[0.9ex]
3419 & 4128 & $\beta$ Cet & \ldots & 170 & 170\rlap{\,$^{a}$} & 169 & 181.1 & 215.0\rlap{\,$^{(13)}$} & 210 & 4.1\rlap{\,$^{(5)}$}\hphantom{\,$^{(11)}$} & 17.1 & HB &  \\
 &  &  &  & $7$ & $^{+8}_{-7}$ & $4$ & $1.1$ &  & $^{+30}_{-24}$ & $0.5$\hphantom{\,$^{(11)}$} & $^{+0.3}_{-0.7}$ &  &  \\[0.9ex]
4906 & 6186 & $\varepsilon$ Psc & \ldots & 255.3 & \ldots & 129.3 & \ldots & 255.3\rlap{\,$^{(12)}$} & 304 & 1.8\rlap{\,$^{(11)}$}\hphantom{\,$^{(11)}$} & 11.08 & HB &  \\
 &  &  &  & $1.8$ &  & $2.1$ &  & $1.8$ & $^{+305}_{-105}$ & $1.0$\hphantom{\,$^{(11)}$} & $^{+0.15}_{-0.21}$ &  &  \\[0.9ex]
16335 & 21552 & $\sigma$ Per & \ldots & 575 & 630 & \ldots & 575 & \ldots & 817 & 2.2\rlap{\,$^{(5)}$}\hphantom{\,$^{(11)}$} & 36.0 & RGB &  \\
 &  &  &  & $5$ & $12$ &  & $5$ &  & $^{+626}_{-251}$ & $1.0$\hphantom{\,$^{(11)}$} & $^{+0.7}_{-0.8}$ &  &  \\[0.9ex]
20252 & 27348 & 54 Per & 4324 & 139 & 139\rlap{\,$^{a}$} & 226 & \ldots & 140\rlap{\,$^{(8)}$} & 338 & 1.4\rlap{\,$^{(7)}$}\hphantom{\,$^{(11)}$} & 9.42 & HB & U, H \\
 &  &  & $240$ & $6$ & $^{+7}_{-6}$ & $9$ &  & $7$ & $^{+155}_{-81}$ & $0.5$\hphantom{\,$^{(11)}$} & $^{+0.11}_{-0.12}$ &  &  \\[0.9ex]
20885 & 28307 & 77 Tau & \ldots & 128 & 128\rlap{\,$^{b}$} & \ldots & 2064 & 140\rlap{\,$^{(2)}$} & 132 & 4.2\rlap{\,$^{(9)}$}\hphantom{\,$^{(11)}$} & 10.99 & HB &  \\
 &  &  &  & $13$ & $13$ &  & $74$ & $18$ & $^{+41}_{-25}$ & $1.0$\hphantom{\,$^{(11)}$} & $^{+0.12}_{-0.12}$ &  &  \\[0.9ex]
21421 & 29139 & $\alpha$ Tau & \ldots & 651 & 651 & \ldots & 591 & \ldots & 517 & 4.3\rlap{\,$^{(9)}$}\hphantom{\,$^{(11)}$} & 44.0 & RGB &  \\
 &  &  &  & $16$ & $16$ &  & $5$ &  & $^{+157}_{-97}$ & $1.0$\hphantom{\,$^{(11)}$} & $^{+0.7}_{-0.7}$ &  &  \\[0.9ex]
23015 & 31398 & $\iota$ Aur & \ldots & 671 & 671 & \ldots & 702 & \ldots & 855 & 7.3\rlap{\,$^{(5)}$}\hphantom{\,$^{(11)}$} & 126 & HB &  \\
 &  &  &  & $8$ & $8$ &  & $11$ &  & $^{+148}_{-117}$ & $1.0$\hphantom{\,$^{(11)}$} & $^{+6}_{-11}$ &  &  \\[0.9ex]
31700 & 47442 & $\nu^3$ CMa & \ldots & 181.6 & 181.6 & \ldots & 196.9 & 183.0\rlap{\,$^{(2)}$} & 347 & 4.3\rlap{\,$^{(5)}$}\hphantom{\,$^{(11)}$} & 29.6 & HB &  \\
 &  &  &  & $1.3$ & $1.3$ &  & $1.4$ &  & $^{+108}_{-68}$ & $1.0$\hphantom{\,$^{(11)}$} & $^{+1.8}_{-1.9}$ &  &  \\[0.9ex]
36962 & 60522 & $\upsilon$ Gem & \ldots & 655 & 645 & \ldots & 655\rlap{\,$^{a}$} & \ldots & 425 & 5.2\rlap{\,$^{(5)}$}\hphantom{\,$^{(11)}$} & 43.7 & RGB & U \\
 &  &  &  & $24$ & $11$ &  & $24$ &  & $^{+102}_{-70}$ & $1.0$\hphantom{\,$^{(11)}$} & $^{+1.8}_{-1.9}$ &  &  \\[0.9ex]
37826 & 62509 & Pollux & \ldots & \ldots & \ldots & \ldots & \ldots & 590.0\rlap{\,$^{(1)}$} & 187 & 2.3\rlap{\,$^{(7)}$}\hphantom{\,$^{(11)}$} & 8.55 & HB &  \\
 &  &  &  &  &  &  &  &  & $^{+45}_{-30}$ & $0.5$\hphantom{\,$^{(11)}$} & $^{+0.03}_{-0.03}$ &  &  \\[0.9ex]
46390 & 81797 & $\alpha$ Hya & \ldots & 525 & 681 & \ldots & 525\rlap{\,$^{a}$} & \ldots & 749 & 4.0\rlap{\,$^{(7)}$}\hphantom{\,$^{(11)}$} & 58.7 & HB & U \\
 &  &  &  & $27$ & $18$ &  & $27$ &  & $^{+140}_{-102}$ & $0.6$\hphantom{\,$^{(11)}$} & $^{+0.6}_{-0.4}$ &  &  \\[0.9ex]
46952 & 82635 & 10 LMi & 4537 & 77.2 & 77.2\rlap{\,$^{a}$} & \ldots & \ldots & 76.6\rlap{\,$^{(8)}$} & 99 & 4.7\rlap{\,$^{(4)}$}\hphantom{\,$^{(11)}$} & 9.18 & HB &  \\
 &  &  & $596$ & $1.0$ & $^{+1.0}_{-1.0}$ &  &  & $1.9$ & $^{+26}_{-17}$ & $1.0$\hphantom{\,$^{(11)}$} & $^{+0.18}_{-0.21}$ &  &  \\[0.9ex]
48356 & 85444 & $\upsilon^1$ Hya & \ldots & 94.2 & 94.2 & 98.5 & \ldots & 91\rlap{\,$^{(8)}$} & 169 & 4.9\rlap{\,$^{(9)}$}\hphantom{\,$^{(11)}$} & 16.4 & HB &  \\
 &  &  &  & $0.3$ & $0.3$ & $1.1$ &  & $5$ & $^{+43}_{-29}$ & $1.0$\hphantom{\,$^{(11)}$} & $^{+0.4}_{-0.4}$ &  &  \\[0.9ex]
60646 & 108225 & 6 CVn & \ldots & 157.3 & \ldots & 157.3 & \ldots & \ldots & 125 & 4.4\rlap{\,$^{(5)}$}\hphantom{\,$^{(11)}$} & 10.89 & HB &  \\
 &  &  &  & $1.9$ &  & $1.9$ &  &  & $^{+36}_{-23}$ & $1.0$\hphantom{\,$^{(11)}$} & $^{+0.17}_{-0.17}$ &  &  \\[0.9ex]
63608 & 113226 & $\varepsilon$ Vir & \ldots & 151.6 & 151.6 & \ldots & \ldots & 120\rlap{\,$^{(8)}$} & 353 & 1.7\rlap{\,$^{(5)}$}\hphantom{\,$^{(11)}$} & 12.15 & HB & U, H \\
 &  &  &  & $0.6$ & $0.6$ &  &  & $8$ & $^{+380}_{-126}$ & $1.0$\hphantom{\,$^{(11)}$} & $^{+0.20}_{-0.10}$ &  &  \\[0.9ex]
64962 & 115659 & $\gamma$ Hya & \ldots & 89 & 89\rlap{\,$^{a}$} & \ldots & \ldots & 142\rlap{\,$^{(8)}$} & 249 & 2.8\rlap{\,$^{(7)}$}\hphantom{\,$^{(11)}$} & 13.6 & HB & U, H \\
 &  &  &  & $9$ & $^{+10}_{-8}$ &  &  & $7$ & $^{+48}_{-35}$ & $0.5$\hphantom{\,$^{(11)}$} & $^{+0.3}_{-0.4}$ &  &  \\[0.9ex]
77512 & 141714 & $\delta$ CrB & 1493 & 59.3 & 59.3\rlap{\,$^{a}$} & 57.22 & 59.91 & 59.0\rlap{\,$^{(2)}$} & 75 & 5.2\rlap{\,$^{(6)}$}\hphantom{\,$^{(11)}$} & 7.67 & HB &  \\
 &  &  & $38$ & $0.5$ & $^{+0.6}_{-0.5}$ & $0.25$ & $0.22$ & $2.0$ & $^{+9}_{-7}$ & $0.6$\hphantom{\,$^{(11)}$} & $^{+0.06}_{-0.07}$ &  &  \\[0.9ex]
80331 & 148387 & $\eta$ Dra & \ldots & 96.8 & 96.8 & \ldots & \ldots & \ldots & 145 & 3.7\rlap{\,$^{(5)}$}\hphantom{\,$^{(11)}$} & 10.48 & HB & U \\
 &  &  &  & $0.6$ & $0.6$ &  &  &  & $^{+54}_{-31}$ & $1.0$\hphantom{\,$^{(11)}$} & $^{+0.25}_{-0.02}$ &  &  \\[0.9ex]
81833 & 150997 & $\eta$ Her & 3053 & 159 & 159\rlap{\,$^{a}$} & 139 & \ldots & 160\rlap{\,$^{(8)}$} & 264 & 1.7\rlap{\,$^{(7)}$}\hphantom{\,$^{(11)}$} & 8.94 & HB &  \\
 &  &  & $137$ & $5$ & $^{+5}_{-4}$ & $3$ &  & $11$ & $^{+96}_{-55}$ & $0.5$\hphantom{\,$^{(11)}$} & $^{+0.02}_{-0.11}$ &  &  \\[0.9ex]
87808 & 163770 & $\theta$ Her & \ldots & 621 & 621 & \ldots & 884 & 166\rlap{\,$^{(8)}$} & 614 & 7.5\rlap{\,$^{(5)}$}\hphantom{\,$^{(11)}$} & 90.5 & HB & U \\
 &  &  &  & $15$ & $15$ &  & $27$ & $6$ & $^{+55}_{-48}$ & $0.6$\hphantom{\,$^{(11)}$} & $^{+1.9}_{-2.2}$ &  &  \\[0.9ex]
87933 & 163993 & $\xi$ Her & 5408 & 120 & 120\rlap{\,$^{a}$} & 121.1 & \ldots & 106\rlap{\,$^{(8)}$} & 115 & 4.5\rlap{\,$^{(5)}$}\hphantom{\,$^{(11)}$} & 10.28 & HB &  \\
 &  &  & $418$ & $3$ & $^{+3}_{-3}$ & $1.0$ &  & $4$ & $^{+17}_{-13}$ & $0.6$\hphantom{\,$^{(11)}$} & $^{+0.06}_{-0.08}$ &  &  \\[0.9ex]
91117 & 171443 & $\alpha$ Sct & \ldots & 267 & 267 & \ldots & 422 & \ldots & 254 & 4.6\rlap{\,$^{(5)}$}\hphantom{\,$^{(11)}$} & 23.1 & HB & U \\
 &  &  &  & $4$ & $4$ &  & $4$ &  & $^{+71}_{-46}$ & $1.0$\hphantom{\,$^{(11)}$} & $^{+1.0}_{-1.2}$ &  &  \\[0.9ex]
96459 & 185351 & \ldots & \ldots & 154 & 154\rlap{\,$^{a}$} & \ldots & 106.4 & 107\rlap{\,$^{(8)}$} & 119 & 2.14\rlap{\,$^{(7)}$}\hphantom{\,$^{(11)}$} & 5.07 & RGB & U \\
 &  &  &  & $22$ & $^{+26}_{-18}$ &  & $0.3$ & $11$ & $^{+14}_{-12}$ & $0.23$\hphantom{\,$^{(11)}$} & $^{+0.05}_{-0.09}$ &  &  \\[0.9ex]
100064 & 192947 & $\alpha^2$ Cap & \ldots & 116 & 116\rlap{\,$^{a}$} & 66.9 & \ldots & 116\rlap{\,$^{(8)}$} & 153 & 2.7\rlap{\,$^{(9)}$}\hphantom{\,$^{(11)}$} & 8.20 & HB &  \\
 &  &  &  & $10$ & $^{+10}_{-10}$ & $1.0$ &  & $9$ & $^{+88}_{-41}$ & $1.0$\hphantom{\,$^{(11)}$} & $^{+0.14}_{-0.15}$ &  &  \\[0.9ex]
100587 & 194317 & 39 Cyg & \ldots & 214.8 & 214.8 & \ldots & \ldots & \ldots & 234 & 4.8\rlap{\,$^{(5)}$}\hphantom{\,$^{(11)}$} & 22.4 & HB & U \\
 &  &  &  & $1.7$ & $1.7$ &  &  &  & $^{+61}_{-40}$ & $1.0$\hphantom{\,$^{(11)}$} & $^{+0.3}_{-0.3}$ &  &  \\[0.9ex]
102422 & 198149 & $\eta$ Cep & 2389 & 546 & 485 & \ldots & 546 & \ldots & 178 & 1.0\rlap{\,$^{(5)}$}\hphantom{\,$^{(11)}$} & 4.01 & RGB & U \\
 &  &  & $130$ & $8$ & $12$ &  & $8$ &  & $^{+248}_{-71}$ & $0.8$\hphantom{\,$^{(11)}$} & $^{+0.05}_{-0.05}$ &  &  \\[0.9ex]
102453 & 197912 & 52 Cyg & \ldots & 238 & 238\rlap{\,$^{a}$} & \ldots & \ldots & 221\rlap{\,$^{(8)}$} & 160 & 4.3\rlap{\,$^{(5)}$}\hphantom{\,$^{(11)}$} & 13.6 & HB & U \\
 &  &  &  & $15$ & $^{+16}_{-14}$ &  &  & $5$ & $^{+49}_{-30}$ & $1.0$\hphantom{\,$^{(11)}$} & $^{+0.4}_{-0.3}$ &  &  \\[0.9ex]
105515 & 203387 & $\iota$ Cap & 4711 & 67.0 & 67.0\rlap{\,$^{a}$} & 68.9 & \ldots & 68\rlap{\,$^{(2)}$} & 127 & 4.4\rlap{\,$^{(7)}$}\hphantom{\,$^{(11)}$} & 10.94 & HB &  \\
 &  &  & $177$ & $0.7$ & $^{+0.7}_{-0.7}$ & $0.5$ &  & $6$ & $^{+15}_{-12}$ & $0.5$\hphantom{\,$^{(11)}$} & $^{+0.21}_{-0.25}$ &  &  \\[0.9ex]
106481 & 205435 & $\rho$ Cyg & \ldots & 120 & 120\rlap{\,$^{a}$} & 109.5 & \ldots & 110\rlap{\,$^{(8)}$} & 102 & 3.9\rlap{\,$^{(9)}$}\hphantom{\,$^{(11)}$} & 7.90 & HB &  \\
 &  &  &  & $14$ & $^{+14}_{-13}$ & $1.7$ &  & $6$ & $^{+35}_{-21}$ & $1.0$\hphantom{\,$^{(11)}$} & $^{+0.03}_{-0.03}$ &  &  \\[0.9ex]
107315 & 206778 & $\varepsilon$ Peg & \ldots & 515 & 585 & 117.7 & 515 & \ldots & 1281 & 8.3\rlap{\,$^{(5)}$}\hphantom{\,$^{(11)}$} & 211 & HB &  \\
 &  &  &  & $7$ & $7$ & $0.6$ & $7$ &  & $^{+303}_{-208}$ & $1.6$\hphantom{\,$^{(11)}$} & $^{+5}_{-6}$ &  &  \\[0.9ex]
117503 & 223460 & OU And & \ldots & 24.45 & \ldots & 24.45 & \ldots & 24.2\rlap{\,$^{(10)}$} & 24.9 & 21.5\rlap{\,$^{(3)}$}\hphantom{\,$^{(11)}$} & 10.6 & HB &  \\
 &  &  &  & $0.08$ &  & $0.08$ &  &  & $^{+2.8}_{-2.3}$ & $2.1$\hphantom{\,$^{(11)}$} & $^{+0.3}_{-0.4}$ &  &  \\[0.9ex]
\end{longtable}
\tablefoot{The errors are given below their respective values. The stellar radii and evolutionary stages are taken from \citet{stock2018}; other references are noted in the superscripts and listed below. For $v \sin i$ values without a published error, we assumed a conservative 1\,km\,s$^{-1}$ error. $P_\mathrm{cyc}$ is the period of the long-term S index modulation. In the last column, U\,=\,uncertain period (see Sect.~\ref{sec:rotation_periods} for the criteria) and H\,=\,possible half (or harmonic) of the rotation period. $^{a}$\,$P_\mathrm{S}$ or $P_\mathrm{rv}$ determined from the GP fit, rather than the periodogram. $^{b}$\,$P_\mathrm{S}$ is the mean and standard deviation of the periods of individual observing seasons.}
\tablebib{(1) \citet{auriere2013}; (2) \citet{choi1995}; (3) \citet{demedeiros1999}; (4) \citet{demedeiros2014}; (5) \citet{hekker2007}; (6) \citet{hojjatpanah2019}; (7) \citet{jofre2015}; (8) \citet{lehtinen2020}; (9) \citet{massarotti2008}; (10) \citet{strassmeier1999}; (11) \citet{takeda2008}; (12) \citet{teng2022}; (13) \citet{tsvetkova2013}}

\newpage

\twocolumn

\section{SOAP spot models}
\label{app:spotmodels}

\begin{table*}
\caption{SOAP models of the stars with a rotation period in the RVs or in both the photometry and the RVs.}
\label{tab:spots}
\centering
\small
\setlength{\tabcolsep}{3.5pt}
\begin{tabular}{lrrrlrrllrrl}
\hline\hline
 & & & & & & & \multicolumn{2}{c}{Single spot} & \multicolumn{3}{c}{Spot and plage} \\
Star & $P$ & $i$ & $\Delta T$ & Fitted to & $K_\mathrm{RV}$ & $\sigma_\mathrm{Hip}$ & $A$ & Pred./obs. & $A_\mathrm{s}$ & $A_\mathrm{p}$ & Pred./obs. \\
 & (d) & ($^\circ$) & (K) & & (m\,s$^{-1}$) & (mmag) & (\%) & & (\%) & (\%) & \\
\hline
$\beta$ Cet & 169.4 / 181.1 & $53^{+11}_{-9}$ & 800 & phot. & 24 & 8.1 & 0.77 & 0.73 (0.67--0.78) & 0.92 & 3.3 & 1.00 (0.94--1.04) \\
$\varepsilon$ Psc$^{a}$ & 258.6 / 255.3 & $36^{+27}_{-14}$ & 800 & phot. & 17$^{b}$ & 5.4 & 0.88 & 0.38 (0.25--0.64) &  &  &  \\
$\delta$ CrB & 57.2 / 59.9 & $52^{+9}_{-7}$ & 600 & phot. & 28 & 9.9 & 1.3 & 1.05 (0.98--1.11) & 1.3 & 0 & 1.05 (0.98--1.11) \\
$\xi$ Her$^{c}$ & 121.1 & $67^{+10}_{-10}$ & 600 & phot. & 5.5 & 12.1 & 1.8 & 5.27 (5.07--5.32) & 1.8 & 0 & 5.27 (5.07--5.32) \\
\hline
$\varepsilon$ Psc & 255.3 & $36^{+27}_{-14}$ & 800 & RV & 17$^{b}$ & 5.4 & 2.3 & 2.47 (2.06--2.81) & 0.50 & 4.8 & 0.30 (0.30--0.32)$^{d}$ \\
$\sigma$ Per & 575.0 & $37^{+20}_{-15}$ & 900 & RV & 127 & 5.7 & 9.5 & 12.32 (10.61--14.02) & 9.4 & 19 & 11.19 (9.71--11.19) \\
$\alpha$ Tau & 591.1 & $65^{+12}_{-11}$ & 900 & RV & 133 & 8.4 & 5.8 & 6.03 (5.95--6.38) & 5.8 & 17 & 4.99 (4.85--5.33) \\
$\iota$ Aur & 701.5 & $51^{+11}_{-9}$ & 900 & RV & 192 & 5.3 & 3.4 & 5.42 (4.93--6.05) & 3.4 & 3.4 & 5.35 (4.87--5.97) \\
$\nu^3$ CMa & 196.9 & $31^{+9}_{-8}$ & 900 & RV & 17 & 5.7 & 0.65 & 0.71 (0.63--0.78) & 0.65 & 0 & 0.71 (0.63--0.78) \\
$\upsilon$ Gem & 654.8 & $74^{+8}_{-9}$ & 900 & RV & 128 & 10.2 & 6.5 & 5.34 (5.34--5.41) & 5.4 & 27 & 2.94 (2.81--2.94) \\
$\alpha$ Hya & 524.7 & $44^{+10}_{-8}$ & 900 & RV & 86 & 6.5 & 3.0 & 3.67 (3.33--4.03) & 2.0 & 22 & 1.02 (0.98--1.02) \\
HIP 96459 & 106.4 & $69^{+10}_{-12}$ & 900 & RV & 7.4 & 5.2 & 0.63 & 0.83 (0.83--0.86) & 0.63 & 0 & 0.83 (0.83--0.86) \\
$\eta$ Cep & 546.1 & $60^{+14}_{-13}$ & 900 & RV & 12 & 3.7 & 5.5 & 10.54 (10.44--10.97) & 0.68 & 3.3 & 1.00 (0.97--1.01) \\
$\varepsilon$ Peg & 515.4 & $24^{+5}_{-5}$ & 900 & RV & 324 & 11.5 & 5.1 & 2.43 (2.26--2.61) & 5.1 & 5.1 & 2.36 (2.22--2.51) \\
$\varepsilon$ Peg$^{e}$ & 1030.9 & $51^{+13}_{-11}$ & 900 & RV & 324 & 11.5 & 3.6 & 1.56 (1.47--1.68) &  &  &  \\
\hline
\end{tabular}
\tablefoot{$P$ is the rotation period of the model (photometric / RV model where they differ), $i$ the median of the inclinations $\sin i = P_\mathrm{rot}/P_\mathrm{max}$ from the draws with $P_\mathrm{rot}<P_\mathrm{max}$ (Sect.~\ref{sec:spotmodel}), with the range between their 16th and 84th percentiles, and $\Delta T$ the spot temperature contrast. Each model is sized so that its standard deviation over one rotation equals that of the best-fitting sine, $K/\sqrt{2}$, in the dataset in which the signal was found (column `Fitted to': photometry or RVs). $K_\mathrm{RV}$ is the semi-amplitude of the best-fitting sine to the RVs and $\sigma_\mathrm{Hip}$ the standard deviation of the \textsc{Hipparcos} photometry. `Pred./obs.' compares the model with the other dataset: for photometry-fitted models, the predicted RV standard deviation over $K_\mathrm{RV}/\sqrt{2}$; for RV-fitted models, the predicted photometric standard deviation over $\sigma_\mathrm{Hip}$, which includes the photometric noise, so that values below unity are consistent with the data. Both models are placed at a latitude of $30^\circ$; $A$ is the size of the single spot as a percentage of the visible hemisphere. In the second model, the spot ($A_\mathrm{s}$) is surrounded by a plage ($A_\mathrm{p}$), with the smallest plage-to-spot area ratio that brings the prediction to the observation, or the one that comes closest; where no plage is needed, it is the single spot. The ranges of Pred./obs. are over the models at the three inclinations; bounding models in which SOAP breaks down are omitted. $^{a}$\,Two spots of this size at opposite longitudes, rotating at twice the photometric period; the RVs are predicted with one such spot at the RV period. $^{b}$\,Lick and HIDES RVs \citep{teng2022} combined. $^{c}$\,No RV signal at this period; the predicted RV standard deviation is comparable to the total scatter of the Lick RVs (25\,m\,s$^{-1}$). $^{d}$\,One active region, fitted jointly to the RV and photometric semi-amplitudes at the rotation period and its half ($\chi^2=4.9$ over the four amplitudes); Pred./obs. as for the RV-fitted models. $^{e}$\,Two spots of this size at opposite longitudes on a star rotating at twice the RV period, the case in which the 515\,d signal is half the rotation period.}
\end{table*}

\begin{figure}[htbp]
    \centering
    \includegraphics[width=\linewidth]{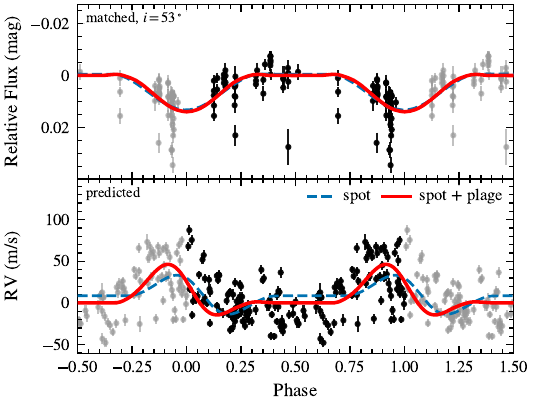}
    \caption{Same as Fig.~\ref{fig:delta_crb_soapmodel} but for $\beta$ Cet. The dashed blue curve shows a single spot of $A=0.77$\,\%, and the red curve a spot of $A_\mathrm{s}=0.92$\,\% surrounded by a plage of $A_\mathrm{p}=3.3$\,\%, with $\Delta T = 800$\,K and $i=53^\circ$.}
    \label{fig:betacet_soapmodel}
\end{figure}

\begin{figure}[htbp]
    \centering
    \includegraphics[width=\linewidth]{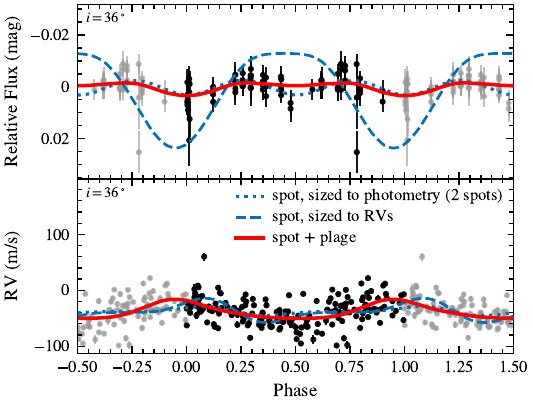}
    \caption{\textsc{Hipparcos} light curve (top) and combined HIDES \citep{teng2022} and Lick RVs (bottom) of $\varepsilon$ Psc, both phase folded at the RV period of 255.3\,d. The dotted curve shows two spots at opposite longitudes ($A=0.88$\,\% each), sized to the photometric signal at 129\,d; the dashed curve one spot ($A=2.3$\,\%) sized to the RVs; and the red curve one spot ($A_\mathrm{s}=0.50$\,\%) surrounded by a plage ($A_\mathrm{p}=4.8$\,\%), fitted to the RV and photometric semi-amplitudes at 255.3 and 128\,d. All are at a latitude of $30^\circ$, with $\Delta T = 800$\,K and $i=36^\circ$.}
    \label{fig:eps_psc_soapmodel}
\end{figure}

\begin{figure}[htbp]
    \centering
    \includegraphics[width=\linewidth]{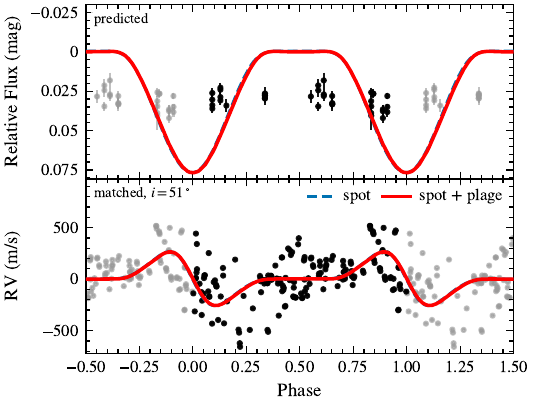}
    \caption{Same as Fig.~\ref{fig:delta_crb_soapmodel} but for $\iota$ Aur, phase folded at the RV period of 702\,d, with the models sized to the RVs: a single spot of $A=3.4$\,\% (dashed blue curve) and a spot of $A_\mathrm{s}=3.4$\,\% surrounded by a plage of $A_\mathrm{p}=3.4$\,\% (red curve), with $\Delta T = 900$\,K and $i=51^\circ$.}
    \label{fig:iotaaur_soapmodel}
\end{figure}
\section{Discussion of individual stars}
\label{app:individual_stars}

\subsection{\texorpdfstring{$\sigma$ Per}{sigma Per} (HIP 16335 / HD 21552)}

The K3\,III giant $\sigma$ Per was reported to host a planetary companion in a $(579.8 \pm 2.4)$\,d orbit by \citet{lee2014}, based on RV measurements from the Bohyunsan Observatory Echelle Spectrograph. \citet{spaeth2025} recently refuted this planet using combined Lick, SONG, and CARMENES data, finding that the RV variations are intrinsically induced, with significant periodicity in the chromatic RV index and bisector span of the CARMENES visual channel near the published planet period.

The S index (49 nights) shows a long-term trend, whose strongest period before detrending is $(5600\pm900)$\,d. After detrending with $P_{\rm cut}=2500$\,d, a single significant peak remains at $(630\pm12)$\,d (FAP between 0.1 and 1\,\%). The Lick RVs show their strongest period at $(575\pm5)$\,d, close to the refuted planet period. The two periods differ by more than their formal uncertainties. The difference is, however, smaller than the width of the S index periodogram peak, which extends from about 570 to 670\,d at half maximum, so the two signals cannot be separated with the HKP data. We adopted the RV period as the rotation period. The maximum rotation period derived from $v \sin i$ is $817_{-251}^{+626}$\,d, placing the 575--630\,d periods well within the plausible range for rotational modulation.

\subsection{\texorpdfstring{$\alpha$ Hya}{alpha Hya} (HIP 46390 / HD 81797)}
\label{sec:alpha_hya}

The RV variations of $\alpha$ Hya were already discussed by \citet{spaeth2025}. They attributed them to an intrinsic stellar origin based on Lick, SONG, and CARMENES observations, finding periods of 524\,d, 306\,d, and 422\,d, respectively. The S index of $\alpha$ Hya consists of only 23 nights. After detrending with $P_{\rm cut}=2500$\,d, the highest periodogram peak is at $(681\pm18)$\,d, with a FAP between 1 and 5\,\%. The Lick RVs show a strong period at $(522\pm6)$\,d. A GP fit to the RVs confirms it at $(525^{+30}_{-24})$\,d with $S_{\mathrm{GP}}=5.7$, in agreement with the 524\,d that \citet{spaeth2025} found in the Lick data, and we adopted this period. The S index periodogram has a secondary peak close to it at 515\,d, which stays below the 5\,\% FAP level. The S index variability, together with the findings from \citet{spaeth2025}, supports the intrinsic origin of the RV variations. Given the sparse S index data and the deviating periods, a possible rotation period of $\approx520$\,d is identified for $\alpha$ Hya, but its value should be treated as uncertain.

\subsection{\texorpdfstring{$\beta$ Oph}{beta Oph} (HIP 86742 / HD 161096)}
\label{sec:beta_oph}

\citet{hatzes1996} found variations at several timescales in $\beta$ Oph using two years of RV data from McDonald Observatory. They attributed the shorter periods to stellar oscillations, and suspected a residual 142\,d period with a semi-amplitude around 30\,m\,s$^{-1}$ to be caused by rotational modulation. The Lick RVs show a similar amplitude of variation, but we did not detect significant periodicity around 140\,d. The only significant periodogram peak is at 424\,d. The S index of $\beta$ Oph consists of only 11 nights. The periodogram shows a peak at $(153.5\pm1.3)$\,d, but it does not reach the 5\,\% FAP level. We therefore do not list a rotation period for $\beta$ Oph, although the S index remains suggestive of the 142\,d period proposed by \citet{hatzes1996}.

\subsection{\texorpdfstring{$\alpha^2$ Cap}{alpha2 Cap} (HIP 100064 / HD 192947)}
\label{sec:alpha2_cap}

The periodogram of the S index data shows several peaks around 100\,d, with the highest power at $(133.2\pm0.3)$\,d. The best-fit GP has a posterior peak at $(116.5 \pm 9.8)$\,d and a significant likelihood over the non-periodic kernel at $S_{\mathrm{GP}}=6.2$. In the \textsc{Hipparcos} photometry, there is a significant peak at $(66.9\pm1.0)$\,d, or about half of the S index period, which could be connected via rotation, considering the maximum rotation period of $(153^{+88}_{-41})$\,d.

\subsection{\texorpdfstring{$\eta$ Cep}{eta Cep} (HIP 102422 / HD 198149)}

The S index data show a dominant periodogram peak at $(485\pm12)$\,d, with a FAP between 1 and 5\,\%, in rough agreement with a significant Lick RV period at $(546\pm8)$\,d. Both periods exceed the upper limit of $178^{+248}_{-71}$\,d, but the $v \sin i$ measurement of $(1.0\pm0.8)$\,km\,s$^{-1}$ is highly uncertain, with line broadening near the spectral resolution limit, rendering $P_\mathrm{max}$ unreliable. A rotation period near 500--550\,d therefore cannot be excluded and we flagged it as a possible but uncertain period.

\subsection{\texorpdfstring{$\iota$ Cap}{iota Cap} (HIP 105515 / HD 203387)}
\label{sec:iota_cap}

For $\iota$ Cap, S index data are available until 2001 in an extended part of the HKP dataset \citep{radick2018a}. It shows long-term activity modulation, leading to a period of about 13\,yr ($4710\pm180$\,d) in the undetrended data.
After detrending with $P_{\rm cut}=1000$\,d, the residuals show a strong periodogram peak at $(67.32 \pm 0.05)$\,d, in agreement with the $(68 \pm 6)$\,d period that \citet{choi1995} found in the shorter version of this dataset. The GP modelling leads to a period of $(67.0 \pm 0.7)$\,d, with $S_{\mathrm{GP}}=36.5$. For the GP fit, the part from the extended dataset was left out, since it shows considerably larger scatter, which degrades the fit.

The \textsc{Hipparcos} photometry shows a significant periodogram peak at $(68.9 \pm 0.5)$\,d. Since the same period appears in both datasets and is below the maximum period of $(127^{+15}_{-12})$\,d, it is likely the star's rotation period.

\subsection{77 Tau (HIP 20885 / HD 28307)}
\label{sec:77_tau}

The S index data of 77 Tau were searched for rotational signals by \citet{choi1995}. They found a period of $(140\pm18)$\,d as the mean over intervals of two consecutive observing seasons. The now available dataset contains four additional observing seasons. After subtracting the long-term trend, the highest-power period is at $(216.9\pm0.9)$\,d, followed by significant peaks at $(179.3\pm1.0)$\,d, $(133.2\pm0.5)$\,d, and $(107.5\pm0.6)$\,d. Analysing each observing season individually, a period of $(128\pm13)$\,d is found. This is the mean and standard deviation over the four seasons in which the period is resolved, which give 112--141\,d. In three further seasons, the variation is at least as long as the season itself, about 180--200\,d, which matches the 179 and 217\,d peaks. Intervals of two consecutive seasons, as used by \citet{choi1995}, resolve this longer variation at $(196\pm4)$\,d and $(186\pm8)$\,d, while three others give 137--151\,d. An average over such intervals therefore mixes the two signals, which may explain the longer period found by \citet{choi1995}.
The best-fit GP model is at $P=(134^{+28}_{-23})$\,d, though with $S_{\mathrm{GP}}=-3.0$, in agreement with the average over individual observing seasons. We adopted the mean of the seasonal periods, $(128\pm13)$\,d, since it separates the two signals and does not rest on a single peak of a multi-peaked periodogram.
In the literature there are two deviating measurements for the projected rotational velocity. \citet{massarotti2008} measured $v \sin i = 4.2$\,km\,s$^{-1}$ without giving an error, while \citet{jofre2015} measured $v \sin i = (1.4 \pm 0.5)$\,km\,s$^{-1}$. Assuming an error of $1$\,km\,s$^{-1}$ for the first measurement, these values lead to an upper limit for the rotation periods of $P = (132^{+41}_{-25})$\,d and $P = (397^{+225}_{-106})$\,d, respectively. A rotation period around 130\,d would be consistent with both $v \sin i$ limits. The instability of this period over different seasons can be explained by the dynamic nature of surface features. The stronger peak at 217\,d is only compatible with the lower $v \sin i$ of \citet{jofre2015}. If it were the rotation period, the 107\,d peak could be its first harmonic, but the 133\,d peak, which recurs in individual seasons and agrees with \citet{choi1995}, could not. We therefore consider a period near 130\,d, as adopted, the more likely rotation period.

\subsection{52 Cyg (HIP 102453 / HD 197912)}
\label{sec:52_cyg}

The S index data show a multiperiodic behaviour with the amplitude of the signal changing drastically, indicating the existence of several activity cycles. After detrending with $P_{\rm cut}=2500$\,d, the periodogram shows significant peaks at 676, 1405, 206, 281, 192, and 150\,d, in the order of prewhitening. Considering the $v \sin i$ of this star leading to an upper limit of $(160^{+49}_{-30})$\,d, the periodogram periods around 150\,d, 192\,d, and 206\,d could be caused by rotational modulation.
The best-fit GP is at a period of $(238^{+16}_{-14})$\,d with $S_{\mathrm{GP}}=6.5$. We adopted it as our best estimate for the rotation period. It lies above the conservative upper limit of 209\,d, but with $1.5\,\sigma$ above the nominal $P_\mathrm{max}$ still within the plausible range.

\subsection{\texorpdfstring{$\gamma$ Hya}{gamma Hya} (HIP 64962 / HD 115659)}
\label{sec:gamma_hya}

The S index data show clear modulation during the individual observing seasons. In the periodograms there are peaks around 150\,d and around 90\,d. The only significant short period is $(99.6\pm0.5)$\,d (FAP between 1 and 5\,\%), next to a long-term signal at 1030\,d. The best-fit GP is at $(89^{+10}_{-8})$\,d, with $S_{\mathrm{GP}}=4.4$, just below the significance threshold, favouring the shorter period.

When analysing the observing seasons individually, there are five seasons with at least 12 nights, all of which show significant periodic variations. Out of these, there are three seasons with a best-fit period between 85\,d and 96\,d, one at $(125\pm16)$\,d, and one in which the variation is at least as long as the season ($\gtrsim146$\,d). If these variations are the result of evolving surface features and rotation, the discrepancy could be caused by decay and migration of spots. At a maximum rotation period of $(249^{+48}_{-35})$\,d, both periods are possible, but the longer period would have to be favoured as the underlying rotation period since shorter-period modulations can be explained by spots at opposite longitudes. We adopted the GP period of $(89^{+10}_{-8})$\,d and flagged it as possibly half the rotation period, which would then be $(178\pm18)$\,d.

\subsection{HIP 96459 / HD 185351}
\label{sec:hip_96459}

The two significant peaks in the S index periodogram of HIP 96459 are at $(141.6\pm0.7)$\,d and $(111.8\pm0.5)$\,d, both with FAPs between 1 and 5\,\%. The Lick RVs show peaks of similar height at $(106.4\pm0.4)$\,d and 149\,d, both close to the 5\,\% FAP level. These two are one-year aliases of each other, so the RVs support a signal near one of the two S index periods, but cannot tell which. The GP fit to the RVs is not significant, but the posterior distribution also shows two peaks, around 110\,d and around 150\,d.
The GP fit to the S index data favours the longer period at $(154^{+26}_{-18})$\,d, but has $S_{\mathrm{GP}}=1.1$, below the significance threshold. We still use it as our best estimate for the underlying period, even though it exceeds the nominal upper limit of $P_{\mathrm{max}}=119^{+14}_{-12}$\,d by $1.5\,\sigma$.

\subsection{\texorpdfstring{$\theta$ Her}{theta Her} (HIP 87808 / HD 163770)}
\label{sec:theta_her}

In the nightly means, detrended with $P_{\rm cut}=1000$\,d, the periodogram shows a strong peak at $(134.4\pm0.3)$\,d and, after prewhitening, two further significant peaks at $(621\pm15)$\,d and $(152.8\pm0.9)$\,d, both with FAPs between 0.1 and 1\,\%. The Lick RVs show periods at $(884\pm27)$\,d and $\approx3900$\,d.
We adopted the 621\,d period, which agrees with the upper limit of $(614^{+55}_{-48})$\,d and would imply a nearly equator-on orientation. The shorter 134 and 153\,d periods would imply an inclination of about $13$--$15^{\circ}$ and, if caused by spots, are more likely to reflect the evolution of individual active regions than rotation. Since the 621\,d period is only the second-strongest signal and is close to the detrending cutoff, it should be regarded as tentative.

\subsection{OU And (HIP 117503 / HD 223460)}
\label{sec:ou_and}

OU And is the fastest rotator in our sample. \citet{strassmeier1999} found OU And to be variable in photometric measurements from October 1996 through June 1997 with a period of 24.2\,d. Periods between 22\,d and 24\,d were found previously by \citet{hopkins1985} and \citet{strassmeier1988} with an amplitude between $0.02$\,mag and $0.06$\,mag. The \textsc{Hipparcos} photometry shows a period of ($24.45\pm0.08$)\,d at an amplitude of 0.008\,mag.

In the MASCARA survey data \citep{burggraaff2018}, there is a light curve available for OU And. The light curve shows strong systematic variations over one night, as was described by \citet{burggraaff2018}. Because the relevant periods are longer than one day, we averaged the data in 1\,d bins, using the standard error of the mean as flux error for each bin and discarding bins with errors above 2.8\,mmag. This removes the noisiest 17\,\% of the nights, mostly those with only a few exposures, but does not change the two dominant periods described below. The periodogram shows two dominant peaks, both with permutation FAPs below 0.1\,\%: a long-term trend around 450\,d, about the length of the light curve, and a second peak around 50\,d. After subtracting the long-term trend, the strongest period remains at ($49.1\pm1.2$)\,d, which is double the period that was previously found, with an amplitude of $(0.0055 \pm 0.0006)$\,mag. After prewhitening, further peaks remain around 30\,d, possibly caused by the lunar cycle, and at $(25.1\pm0.5)$\,d, close to the \textsc{Hipparcos} period. Neither reaches the 5\,\% permutation FAP level (analytic FAPs of 0.006 and 0.2\,\%), so the 25\,d signal is at most suggestive in these data.
The rotational velocity of $v \sin i=(21.5\pm2.1)$\,km\,s$^{-1}$ measured by \citet{demedeiros1999} leads to a maximum period of $(24.9^{+2.8}_{-2.3})$\,d, supporting the shorter period as the true rotation period. \citet{jonsson2020} reported a similar $v \sin i$ of 22.4\,km\,s$^{-1}$. However, it remains unclear how to reconcile this with the 49\,d period observed by MASCARA. While measuring half the true period can be easily explained by a spot configuration, there is no known phenomenon that leads to measuring double the true value.

The 22 Lick RVs of OU And, taken between 2004 and 2011, scatter by 85\,m\,s$^{-1}$, well above their median uncertainty of 32\,m\,s$^{-1}$. Their highest periodogram peaks, at 22.3 and 23.7\,d, lie within the range of published photometric periods, but with a permutation FAP of about 20\,\% they are consistent with noise. A single spot sized to the \textsc{Hipparcos} signal ($A\approx1$\,\%, $i=68^\circ$) predicts an RV scatter of about 60\,m\,s$^{-1}$, so the RVs are consistent with spot modulation, but we do not claim a detection.

There are only seven S index measurements for OU And in the S index data, taken on two nights, which precludes variability analysis. The median of the S index measurements is 0.89, indicating a high level of magnetic activity, consistent with the fast rotation of the star.

\subsection{\texorpdfstring{$\varepsilon$ Psc}{Eps Psc} (HIP 4906 / HD 6186)}
\label{sec:eps_psc}

\citet{teng2022} found variations with $P=255.3^{+2.1}_{-1.4}$\,d and $K=20.6^{+2.1}_{-2.9}$\,m\,s$^{-1}$ in their RV measurements of $\varepsilon$ Psc and attributed them to a planet candidate. Based on a correlation of the RVs with the bisector inverse span (BIS), which is an indicator for line-profile asymmetry, and a similar period in the BIS, they considered intrinsic stellar activity a possible alternative explanation. They state that they did not find significant periodicity in the \textsc{Hipparcos} photometry, ruling out spot modulation.

There is, however, a periodogram peak at $(129\pm2)$\,d in the \textsc{Hipparcos} photometry. With the permutation method, we determined a FAP below $0.1\,\%$ (analytic FAP $0.007\,\%$). Since this period is about half of the RV period, our interpretation is that the 255\,d RV signal traces the rotation period, while the photometric modulation appears at half this value, caused by two opposite active longitudes. Since the \textsc{Hipparcos} data and the RV data are separated by a decade, the spot configurations sampled by the two datasets need not be the same. Figure~\ref{fig:eps_psc_soapmodel} shows that two opposite spots that recreate the photometric amplitude predict only two fifths of the RV scatter, while a single spot sized to the RVs predicts 2.5 times the photometric scatter (Table~\ref{tab:spots}). A single active region, a spot surrounded by a plage with about nine times its area, approximately reproduces the RV semi-amplitude at the rotation period as well as the photometric semi-amplitudes at the rotation period and its half (Fig.~\ref{fig:eps_psc_soapmodel}). The plage shifts the RVs through its suppressed convective blueshift with little photometric signal, and since its brightening is strongest near the limb and the dimming by the spot near disc centre, the region also produces a photometric signal at half the rotation period, comparable to that at the rotation period itself.
Using the $v \sin i$ from \citet{takeda2008} leads to an upper limit on the rotation period of $(304^{+305}_{-105})$\,d, assuming an uncertainty of 1\,km\,s$^{-1}$ as for the other measurements without a published error.

\begin{figure}[ht]
    \centering
    \includegraphics[width=\linewidth]{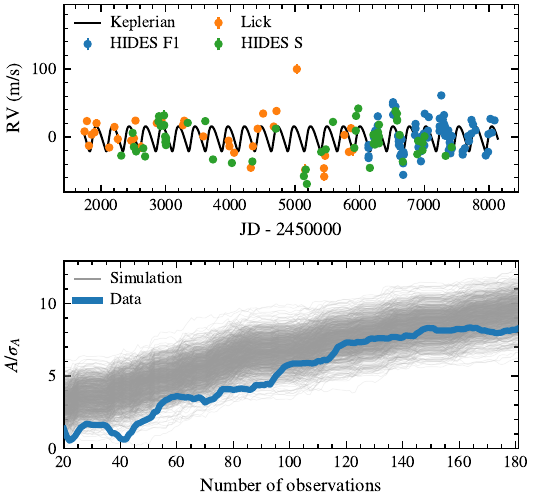}
    \caption{Stacked BGLS analysis of the combined radial velocities of $\varepsilon$ Psc. \textit{Top:} Lick, HIDES-F1, and HIDES-S RVs together with the best-fit Keplerian model at $255$\,d (black line). \textit{Bottom:} Significance, $A/\sigma_A$, of the RV signal, evaluated at the best-fit period, as a function of the number of observations added in time order (blue). The grey lines show the same quantity for $10^3$ realisations of the Keplerian model sampled at the observing epochs with added white noise equal to the RV jitter.}
    \label{fig:eps_psc_sbgls}
\end{figure}

To assess whether the RV signal is coherent or evolving, we combined our Lick RVs with the HIDES-F1 and HIDES-S measurements of \citet{teng2022} and computed a stacked Bayesian generalised Lomb--Scargle (BGLS) periodogram \citep{mortier2017}, tracking the significance, $A/\sigma_A$, of the sinusoid at the best-fit period as observations are added in time, following the approach of \citet{spaeth2025}. The grey lines in Fig.~\ref{fig:eps_psc_sbgls} show $10^3$ realisations of the best-fit Keplerian model, sampled at the observing epochs with added white noise at the level of the stellar jitter. The significance of the real data grows more slowly than in the simulations, especially in the beginning, which supports that it is not a stable Keplerian signal.

Taken together, the photometric period at half the RV period, the approximate consistency of the photometric and RV amplitudes with a spot surrounded by a plage, and the agreement of the 255\,d period with the $v \sin i$-based upper limit lead us to attribute the RV variations to rotational modulation rather than to a planetary companion. We therefore adopted the RV period of \citet{teng2022} as the rotation period of $\varepsilon$ Psc.

\end{appendix}
\end{document}